\documentclass[twocolumn,astrosymb]{aastex631}

\graphicspath{{./}{figures/}}

\newcommand{\pc}{\,\mathrm{pc}}
\newcommand{\kpc}{\,\mathrm{kpc}}
\newcommand{\Msun}{\,\mathrm{M}_\odot}
\newcommand{\kms}{\,\mathrm{km}\,\mathrm{s}^{-1}}
\newcommand{\cc}{\,\mathrm{cm}^{-3}}
\newcommand{\K}{\,\mathrm{K}}
\newcommand{\yr}{\,\mathrm{yr}}
\newcommand{\Myr}{\,\mathrm{Myr}}
\defcitealias{Schneider20}{Schneider20}

\begin{document}

\title{CGOLS V: Disk-wide Stellar Feedback and Observational Implications of the Cholla Galactic Wind Model}

\correspondingauthor{Evan Schneider}
\email{eschneider@pitt.edu}

\author[0000-0001-9735-7484]{Evan E. Schneider}
\affil{Department of Physics \& Astronomy and Pitt PACC, University of Pittsburgh, 100 Allen Hall, 3941 O'Hara Street, Pittsburgh, PA 15260, USA}
\author[0000-0002-2491-8700]{S. Alwin Mao}
\affil{Department of Physics \& Astronomy and Pitt PACC, University of Pittsburgh, 100 Allen Hall, 3941 O'Hara Street, Pittsburgh, PA 15260, USA}

\begin{abstract}
We present the fifth simulation in the CGOLS project -- a set of isolated starburst galaxy simulations modeled over large scales ($10\kpc$) at uniformly high resolution ($\Delta x \approx 5\pc$). Supernova feedback in this simulation is implemented as a disk-wide distribution of clusters, and we assess the impact of this geometry on several features of the resulting outflow, including radial profiles of various phases; mass, momentum, and energy outflow rates; covering fraction of cool gas; mock absorption-line spectra; and X-ray surface brightness. In general, we find that the outflow generated by this model is cooler, slower, and contains more mass in the cool phase than a more centrally concentrated outflow driven by a similar number of supernovae. In addition, the energy loading factors in the hot phase are an order-of-magnitude lower, indicating much larger losses due to radiative cooling in the outflow. However, coupling between the hot and cool phases is more efficient than in the nuclear burst case, with almost 50\% of the total outflowing energy flux carried by the cool phase at a radial distance of 5 kpc. These physical differences have corresponding signatures in observable quantities: the covering fraction of cool gas is much larger, and there is greater evidence of absorption in low and intermediate ionization-energy lines. Taken together, our simulations indicate that centrally-concentrated starbursts are more effective at driving hot, low-density outflows that will expand far into the halo, while galaxy-wide bursts may be more effective at removing cool gas from the disk.

\end{abstract}

\section{Introduction}\label{sec:intro}

Galactic winds, once thought to be an anomalous feature of ``peculiar" galaxies \citep{Lynds63}, have in the last several decades come to be recognized as a key process in galaxy evolution \citep[][and references therein]{Somerville15, Naab17}. By transporting mass, metals, and energy out of galaxies, winds can alter or reverse the effects of cosmic accretion onto halos and explain many observed properties of the galaxy population, including low cosmic star formation efficiency, long gas depletion times, the mass-metallicity relation, metal absorption in the intergalactic medium (IGM), and more \citep[e.g.][]{Larson74, Dekel86, Navarro93, Ellison00, Tremonti04, Erb06, Steidel10}. Given their perceived importance, a vast theoretical effort has been made in recent years to better understand the mechanisms by which galaxies drive winds \citep[][and references therein]{Heckman17}.

A primary tool in this effort are numerical simulations. On the largest scales, cosmological simulations model populations of galaxies, and have had increasing success in recent years in reproducing the observed galaxy population across a range of redshifts \citep{Schaye15, Nelson17, Pillepich18b, Dave19}. This success is in large part due to their adoption of various ``feedback prescriptions" which use physically motivated relationships between galaxy properties like star formation rate and circular velocity in order to eject mass from galaxies \citep[][etc.]{Springel03, Oppenheimer06, Christensen16, Pillepich18a}. While these prescriptions have largely succeeded in that they result in galaxies with appropriate stellar masses and sizes, they vary widely in the precise prescriptions that are chosen, with very different choices for mass ejection, wind speeds, energy injection from supernovae, and more. In addition, most prescriptions implement supernova-driven winds as gas in a single phase with a single velocity \citep[though see][]{Huang22}, potentially missing the impact of high specific energy outflows \citep{Smith21}. Thus, while the universal adoption of such prescriptions is a strong argument for the important role that winds play in galaxy evolution, these large-scale approaches cannot explicitly predict properties of winds on small scales, nor elucidate their driving mechanisms.

As a result, much work has gone into modeling winds on smaller scales and at higher resolution. Ranging from cosmological zooms to volumes that capture only a patch of a single galaxy's interstellar medium (ISM), these simulations attempt to directly resolve the physics that couples supernovae to wind driving \citep{Walch15, Martizzi16, Li17, Kim17, Hopkins18, Fielding18, Kim20a, Martizzi20, Vijayan23}. A major goal is to connect the properties of the resulting winds to local or global galaxy properties such as gas surface density, star formation rate density, galaxy circular velocity, etc. This approach has also seen increasing success in the last decade, with a number of simulations now capable of generating multiphase outflows that show at least broad agreement with observations of winds in the local Universe \citep{Li20}. Nevertheless, a disconnect often exists between the relatively low mass outflow rates measured in the highest resolution simulations and those required by cosmological simulations to reproduce the observed galaxy properties \citep{Pandya21}.

One reason for this disconnect may be unresolved physics. A number of high resolution idealized studies of cool clouds embedded in hot background winds have been carried out which demonstrate that the physics of multiphase winds is not as simple as most cosmological models assume \citep{Cooper09, Scannapieco15, Schneider15, Banda-Barragan16, Schneider17, Abruzzo22}. Depending on the cloud properties and wind conditions, cool clouds in hot outflows can either be destroyed as they are carried out, further mass-loading the hot phase, or grow in mass as hot gas condenses out \citep{Armillotta16, Gronke18, Sparre20, Gronke20, Kanjilal21}. While these studies have proven extremely instructive, they also cannot tell the full story, since real outflows contain populations of many clouds with a spectrum of masses and sizes. This has lead other authors to develop analytic approaches that attempt to capture the range of mass, momentum, and energy transfer that can happen between phases, in order to develop better prescriptions for the next generation of cosmological simulations \citep{Thompson16, Nguyen21, Fielding22}.

A final challenge that each of these theoretical approaches must meet is confrontation with observations themselves. Particularly in the local Universe, the samples of galaxies with observed outflows and measured wind properties has grown immensely in the last decade, especially when considering the cool ionized phase \citep[][etc.]{Martin12, Rubin14, Bordoloi14, Heckman15, Chisholm17, Sugahara17, Xu22, Perrotta23, McPherson23}. Given that this phase has often been assumed to carry out most of the mass, pinning down the column densities and kinematics of outflows in this phase and relating them back to global galaxy properties is a key goal of many studies \citep[][and references therein]{Veilleux20}. In principle, these observations should inform simulations of multiphase outflows, and vice versa. However, connecting the simulation data to the observations is not trivial, particularly close to galaxies where the ionizing photon background is dominated by local sources and much of the outflowing gas is photoionized \citep{Chisholm16}. In addition, many of the commonly observed emission and absorption lines are resonantly scattered, requiring complex radiative transfer models to piece together the full physical picture \citep{Prochaska11}. While nascent efforts in this area exist, a lack of sufficiently high resolution detailed models spanning a range of galaxy properties remains a challenge \citep{Smith22, Carr23}.

Into this broad effort we bring CGOLS, the Cholla Galactic OutfLow Simulation suite \citep{Schneider18a}. The general goal with these simulations is to model supernova driven multiphase outflows across the scale of an entire galaxy ($\sim10\kpc$), but with high enough resolution to capture details affecting the evolution of individual cool clouds, including the transfer of mass, momentum, and energy between phases. With a typical resolution of $\Delta x \approx 5\pc$, the outflow rates measured in CGOLS can be compared directly to ``tall-box" simulations that focus on a single patch of the ISM. Meanwhile, the global properties of the wind are captured on larger scales and can be compared to global galaxy properties and measurements made in cosmological zooms, as well as assumptions made in feedback prescriptions. Finally, the outflows generated in CGOLS are highly complex and structured, making them ideal to compare to a variety of observations, and a good starting point for more complex radiative transfer modeling.

In \cite{Schneider20} (hereafter \citetalias{Schneider20}), we presented CGOLS IV, a simulation of an M82-like starburst galaxy with a clustered stellar feedback prescription. In this paper, as a followup to that work, we analyze a similar simulation, CGOLS V, which has an identical structure but a more extended spatial distribution of clusters (as described in Section \ref{sec:simulations}). We analyze the resulting outflow properties in Sections \ref{sec:profiles} and \ref{sec:fluxes}, and compare the results to the CGOLS IV model in Section \ref{sec:comparison}. In Section \ref{sec:observables}, we present mock column density maps and spectra. Finally, we discuss our results in the context of other simulations of outflows in Section \ref{sec:sim_compare} and observations in Section \ref{sec:obs_compare}.

\section{The Distributed Cluster Simulation}
\label{sec:simulations}

The following section describes the simulation setup for CGOLS V - the distributed cluster model. We begin with a brief description of the initial conditions, followed by a more detailed description of the new cluster feedback implementation. All simulations were run using the \texttt{Cholla} hydrodynamics code \citep{Schneider15} with piecewise parabolic reconstruction, the HLLC Riemann solver, the Van Leer integrator, and cooling prescribed by a parabolic fit to a solar metallicity collisional ionization equilibrium (CIE) cooling curve generated using \texttt{Cloudy} \citep{Ferland13}. We employ diode boundaries on all faces, which allow mass to exit but not flow into the volume. The exact configuration of Cholla used to run the simulations described in this paper can be found here\footnote{https://github.com/alwinm/cholla/tree/foe}.

\subsection{Initial Conditions}

The initial conditions for the CGOLS V model are identical to those used in the previous CGOLS simulations, and we refer the reader to \cite{Schneider18a} for a more detailed description. In brief, the simulations are run in a box with dimensions $L_x = 10\kpc$, $L_y = 10\kpc$, $L_z = 20\kpc$. The simulations are run on a fixed grid, with $n_x = 2048$, $n_y = 2048$, $n_z = 4096$ cells in each dimension, for a constant physical resolution of $\Delta \approx 4.9\pc$. The grid is initialized with a rotating exponential gas disk in vertical hydrostatic equilibrium with properties modeled after the nearby starburst galaxy M82: gas mass $M_\mathrm{disk, gas} = 2.5\times10^9\Msun$ and gas scale radius $R_\mathrm{disk, gas} = 1.6\kpc$ \citep{Greco12}. The disk scale height and circular velocity are calculated including a static gravitational potential consisting of a combination of a Miyamoto-Nagai stellar disk with mass $M_\mathrm{disk, stars} = 10^{10}\Msun$ and scale radius $R_\mathrm{disk, stars} = 0.8\kpc$ \citep{Mayya09}, and an NFW halo with mass $M_\mathrm{halo} = 5\times10^{10}\Msun$, scale radius $R_\mathrm{halo} = 5.3\kpc$, and concentration $c = 10$. This results in a peak circular velocity of $v_\mathrm{circ} \approx 130\kms$ and a peak midplane number density of $n \approx 200\cc$. We also include an adiabatic hot halo with number density $n \approx 10^{-3}\cc$ and temperature $T \approx 2\times 10^6\K$, but note that this halo is blown out of the simulation volume by the feedback-generated outflow and is no longer present at the times we analyze the simulation in Section \ref{sec:results}.

\begin{figure}[t!]
\centering
\includegraphics[width=\linewidth]{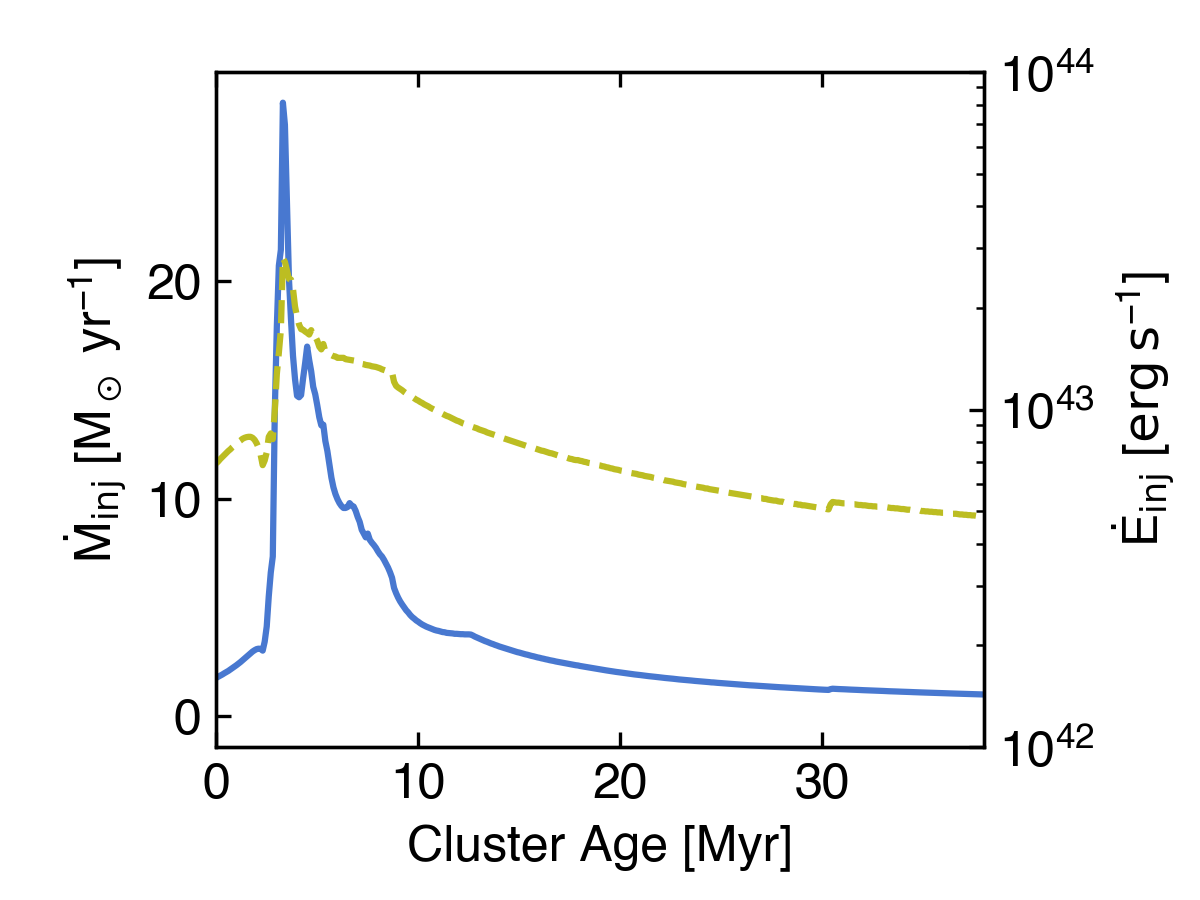}
\caption{Mass (blue solid line) and energy (yellow dashed line) injection rates as a function of time for a single cluster, normalized to the total star formation rate, $\dot{M}_\mathrm{SFR} = 20\Msun\yr^{-1}$.}
\label{fig:S99rates}
\end{figure}

\subsection{Cluster Feedback}\label{sec:feedback}

As in our previous simulations, feedback in the CGOLS V simulation is prescribed via an assumed star-formation rate and associated mass and energy injection from ``star clusters" placed within the simulation volume. Like the CGOLS IV model, these clusters consist of spherical regions with radii of $R_\mathrm{cl} = 30\pc$ into which mass and thermal energy are deposited over time as a function of the cluster age. However, unlike in our previous models, where all clusters had the same mass, here we use a cluster mass function with a PDF $\propto M_\mathrm{cl}^{-2}$, similar to that observed for the nuclear star clusters in M82 and other nearby star-forming galaxies \citep{McCrady07, Mayya08}. We set low and high cutoff masses of $10^4\Msun$ and $5\times10^6\Msun$, respectively. Also in contrast with our previous work, we distribute the clusters broadly throughout the disk, rather than only in the center. Each cluster is assigned a radial location such that the integrated surface density distribution of clusters follows the surface density distribution of the stellar disk, and the total number of clusters as a function of cylindrical radius is proportional to:
\begin{equation}
N_\mathrm{cl} \propto R \, e^{- R / 1.0\,\mathrm{kpc}},
\end{equation}
out to a maximum radius of 4.5 kpc. Thus, the radial distribution of clusters peaks at a scale radius of 1.0 kpc, though the integrated surface density of star formation is highest at the center. Azimuthal locations are chosen randomly, and small ($<10\pc$) random offsets in $z$ are also included in the cluster initial positions. We note that although the aggregate cluster distribution is chosen to follow the disk surface density, \textit{individual} cluster masses are not correlated with local surface density. The list of cluster masses and positions is generated prior to running the simulation and is identical for simulations of different resolutions. 

\begin{figure*}[t!]
\centering
\includegraphics[width=0.95\linewidth]{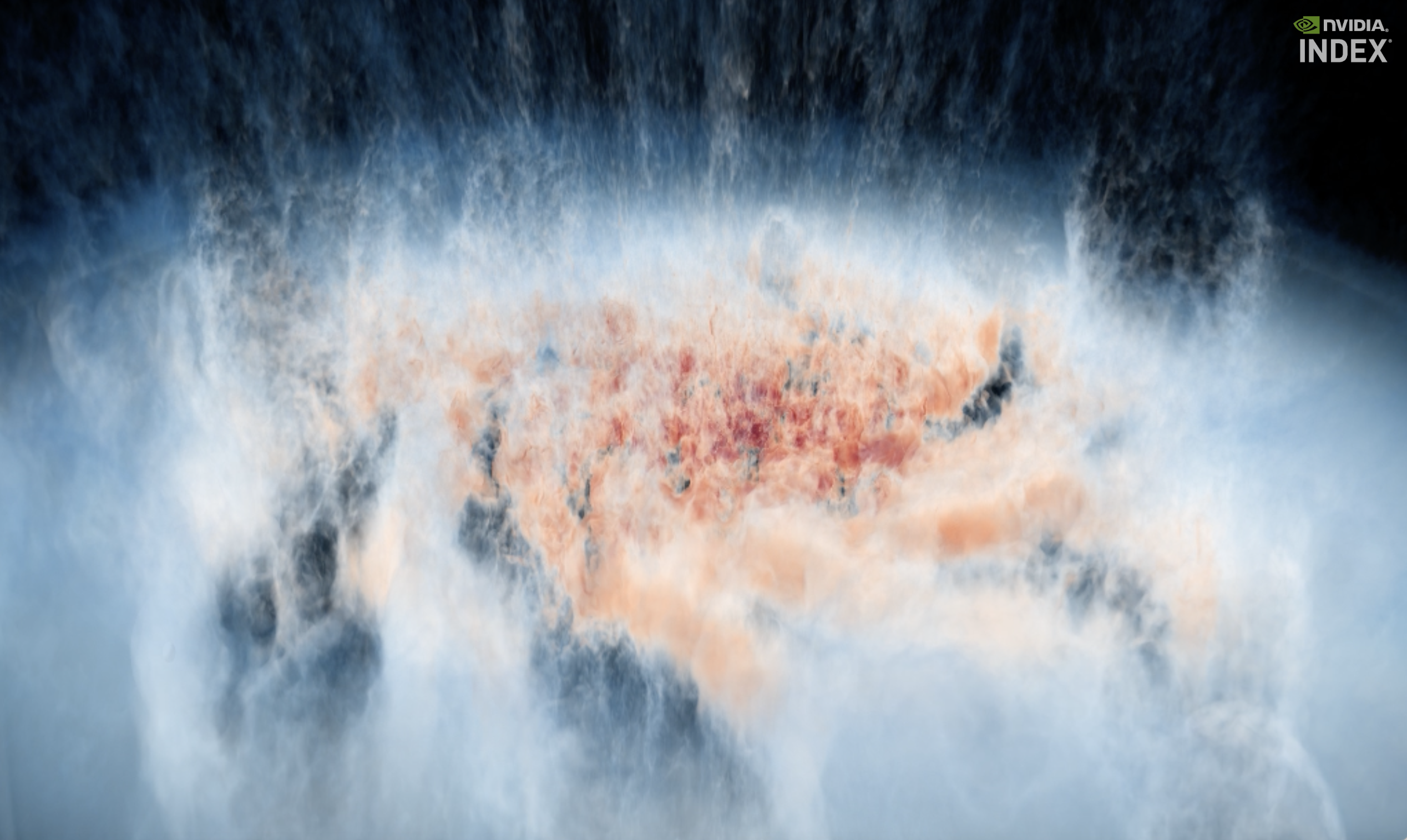}
\caption{Zoomed-in rendering of the density field above the disk, 30 million years after the start of cluster feedback. \label{fig:outflow}}
\end{figure*}

Once the simulation has begun, we start to populate the disk with clusters, which turn on at a rate set by our assumed star formation rate of $20\Msun\,\mathrm{yr}^{-1}$. At every hydro time step we check the cumulative ``stellar mass" generated thus far by summing the mass in existing clusters, and add new clusters to the simulation in order from the list, until the cumulative stellar mass is consistent with the assumed star formation rate. In practice, this means that many time steps may  elapse without a new cluster turning on, particularly if the last cluster added was massive. Aside from this stochasticity, we maintain a constant star formation rate for the duration of the simulation, so new clusters are continually being added.

After turning on, clusters  rotate with the disk according to the circular velocity at their radius. While clusters are on, they continuously deposit mass and thermal energy within their spherical volume according to the yields from a Starburst99 ``single burst" model, scaled appropriately for the cluster mass \citep{Leitherer99}. Figure \ref{fig:S99rates} shows the injection rates for a single cluster, normalized to the overall star formation rate. In terms of ``mass loading" and ``energy loading", on average these rates correspond to $\eta_m = 0.175$ and $\eta_E = 1.25$ (see Section 3.3 for more details). This injection mechanism is similar to that used in the CGOLS IV model, but clusters live longer (40 Myr vs 10 Myr) and inject slightly more mass and energy over longer timescales. As in our previous work, all injected mass is also given a ``color", $s = \rho_s / \rho$, where $\rho_s$ is the color density that is tracked along with the other hydrodynamic variables, and $s$ is a passive scalar variable. The value of the scalar in the simulation volume ranges from 0, for gas that was present in the initial conditions, to 1, for gas that was injected by a cluster.

\section{Results}\label{sec:results}

\subsection{Overview}


After turning on the cluster feedback model described in Section~\ref{sec:feedback}, we evolve the simulation forward in time for $30\Myr$. Unless otherwise specified, our results will focus on the $30\Myr$ simulation snapshot, in order to compare with previous CGOLS models that were analyzed after the same amount of time. A zoomed-in view of the density distribution in the disk is shown in Figure~\ref{fig:outflow}. Here, darker red colors indicate the highest densities, peach is intermediate, and light blue shows more diffuse gas. The least dense gas, corresponding to the volume-filling hot phase of the outflow, is transparent. As Figure \ref{fig:outflow} shows, after $30\Myr$ much of the disk is disturbed by the cluster feedback, and there is outflowing gas at all radii. Higher density clumps tend to be closer to the disk, while more diffuse gas extends up to larger heights above the plane.

\begin{figure*}[ht!]
    \centering
    \includegraphics[width=0.3\linewidth]{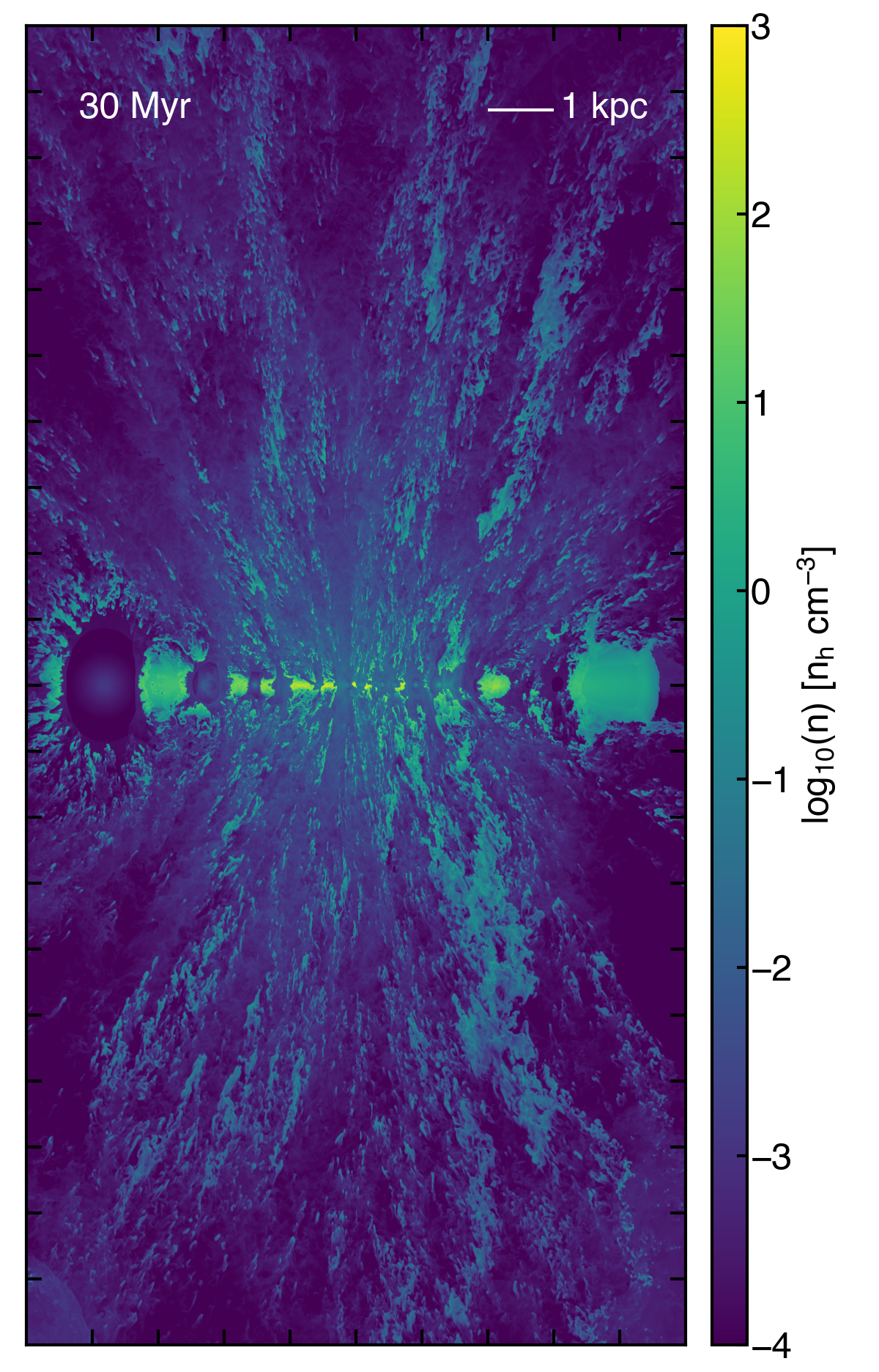}
    \includegraphics[width=0.3\linewidth]{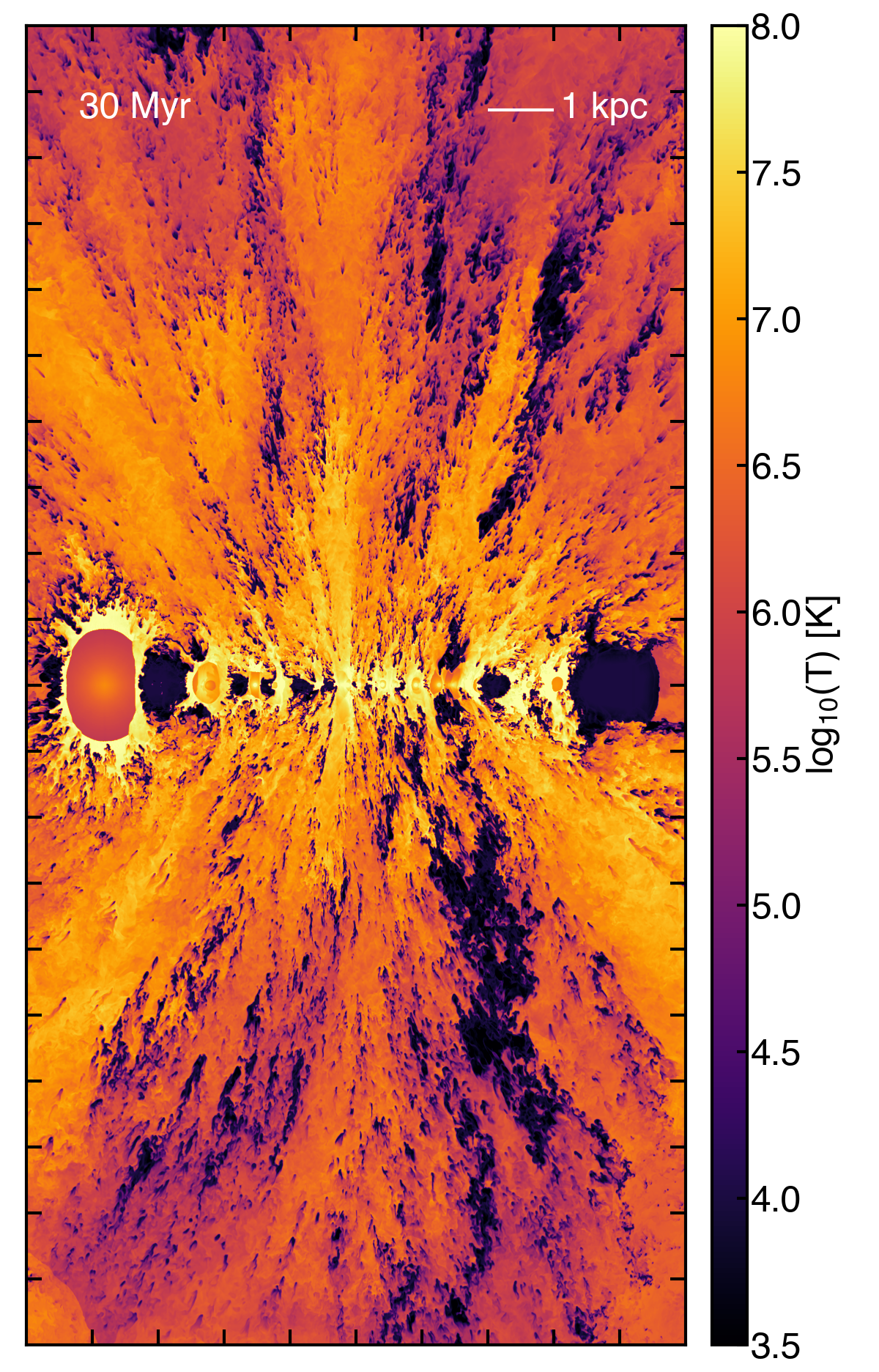}
    \includegraphics[width=0.3\linewidth]{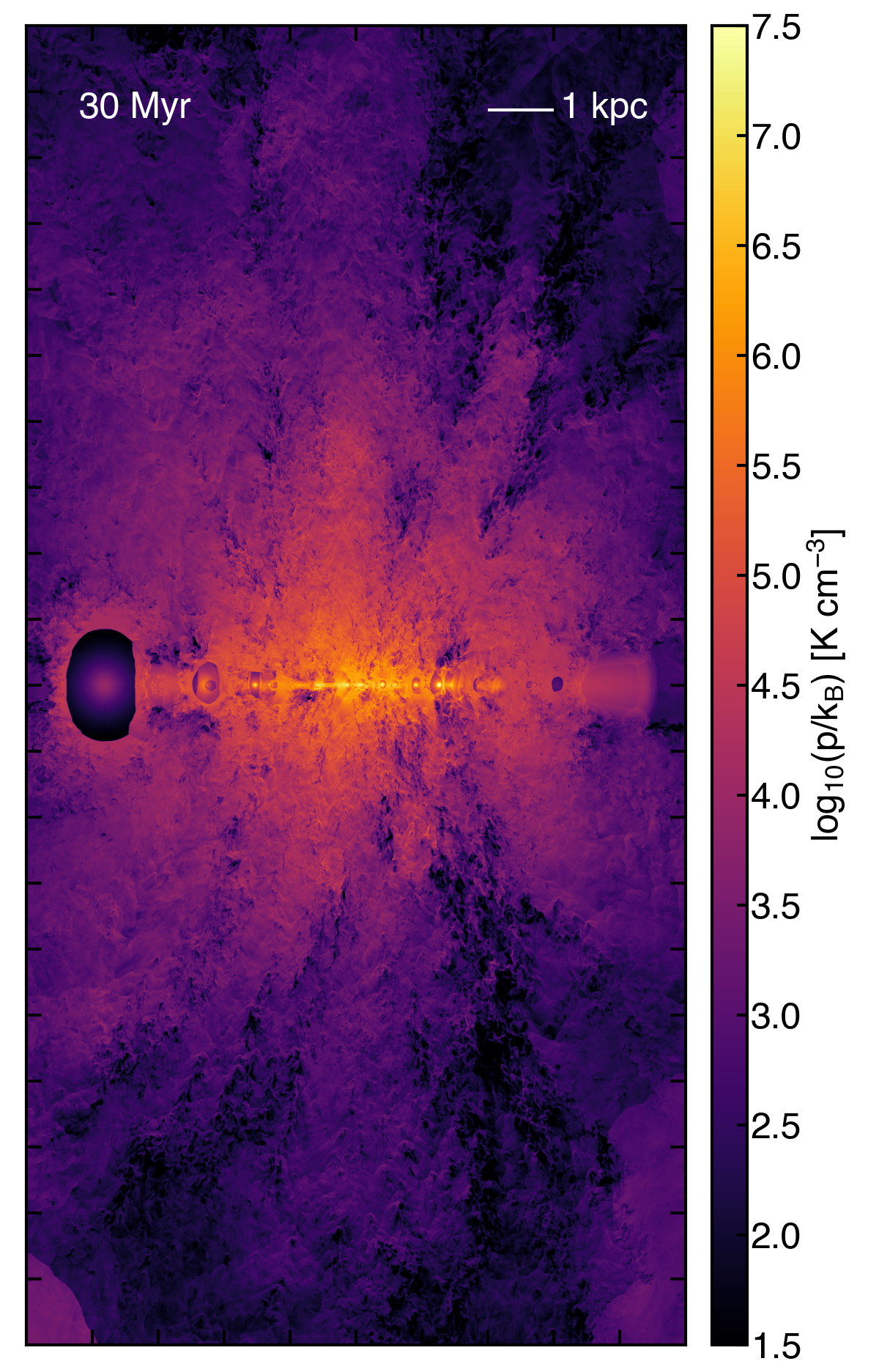}    
    \includegraphics[width=0.3\linewidth]{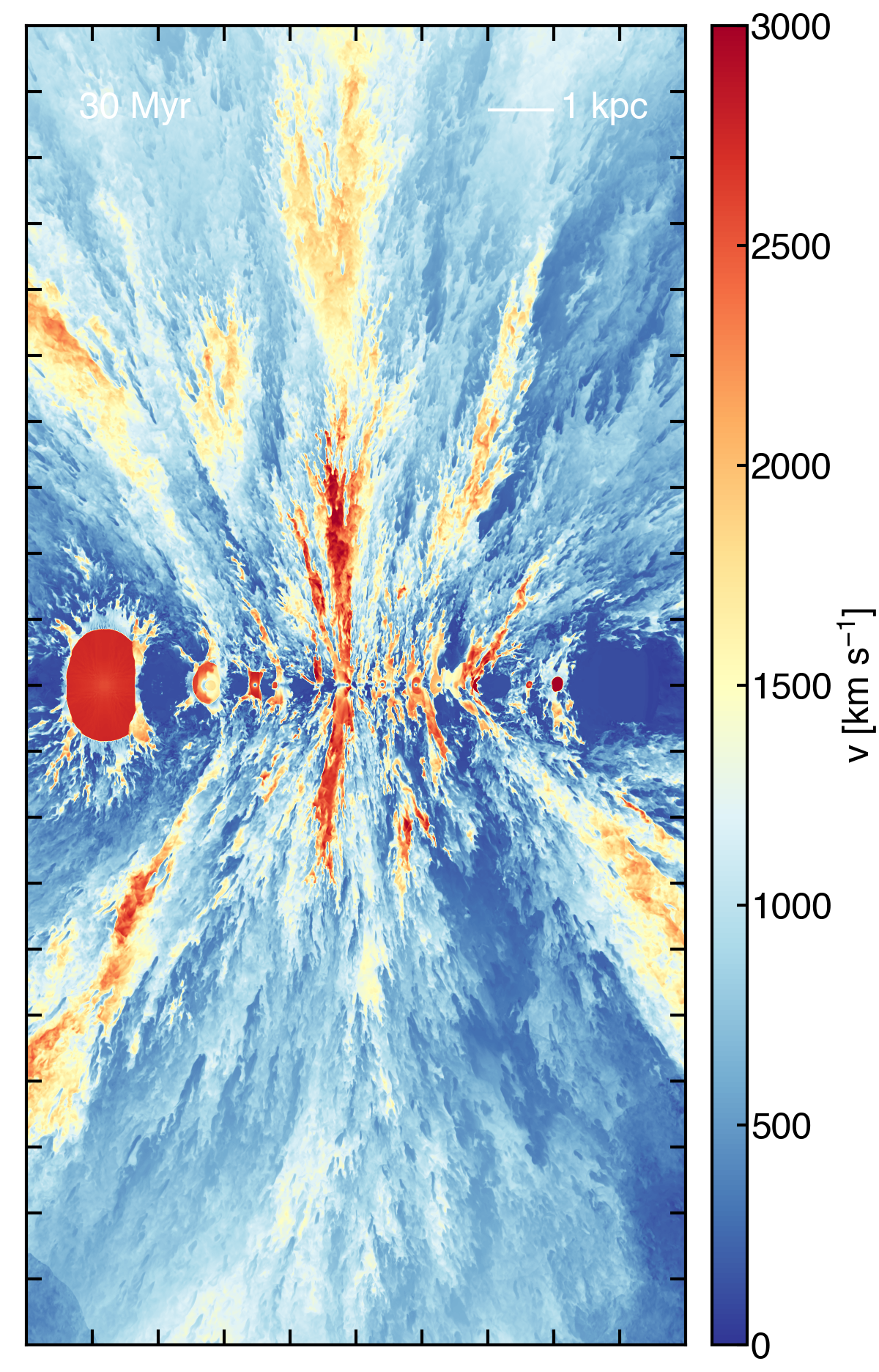}
    \includegraphics[width=0.3\linewidth]{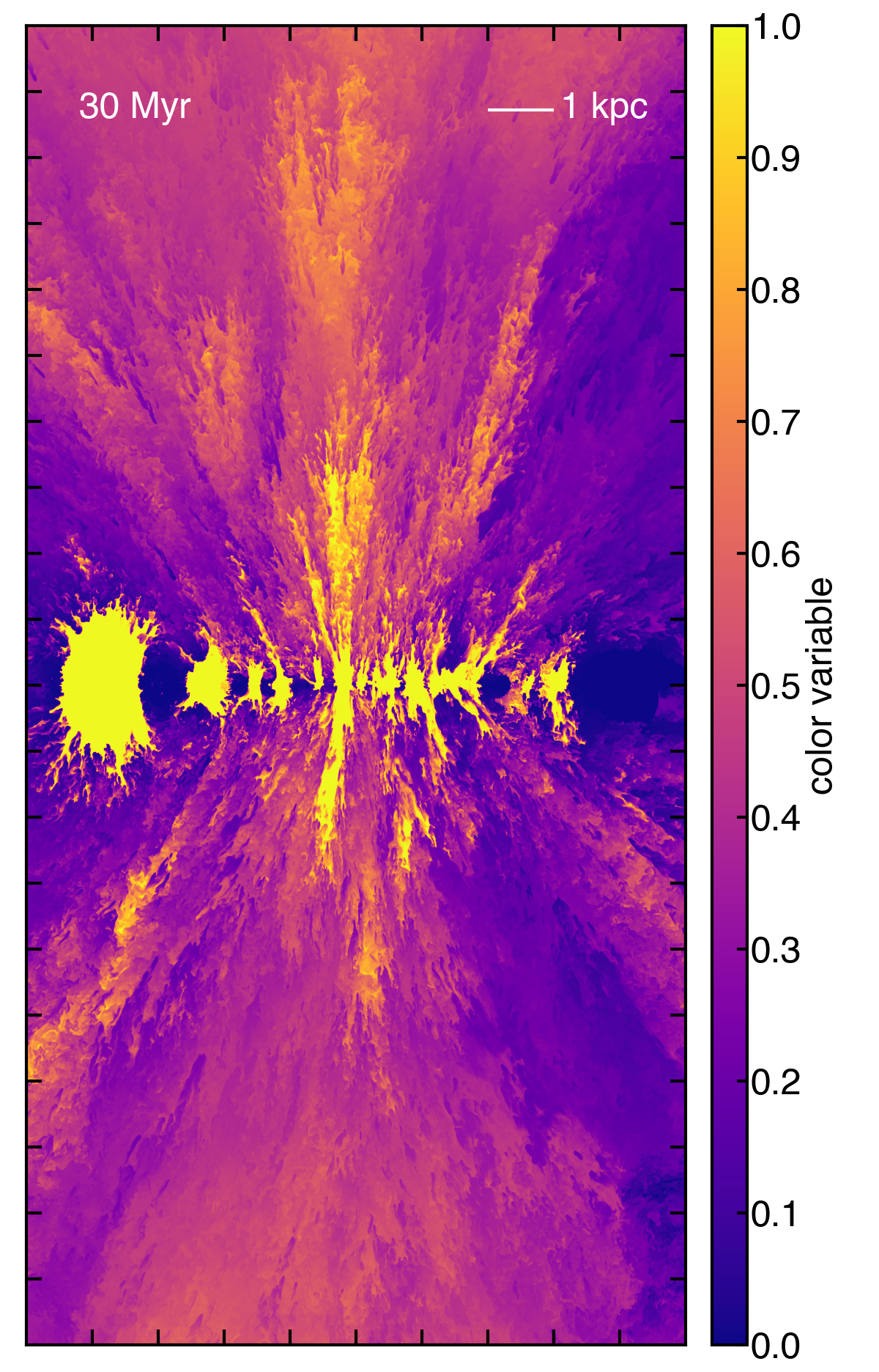}

    \caption{Slices through the $y$-midplane of the simulation showing gas number density, temperature, pressure, velocity, and scalar color, 30 million years after the start of feedback.}
    \label{fig:slices}
\end{figure*}

These features can be seen more quantitatively in Figure~\ref{fig:slices}, which shows $x-z$ slices through the midplane of number density, temperature, pressure, velocity, and color. Here, we see that as in previous CGOLS models, the outflow is characterized by high density cool clouds embedded in a lower density hot, volume-filling phase. Velocity and temperature are correlated, with hotter regions of the outflow traveling at higher speeds. While the cool gas does attain high velocities, it does not tend to exceed $\sim1000\kms$. Cool gas is distributed throughout the simulation volume, and exists out to $10\kpc$, the top of the volume contained in the simulation.

In both Figures \ref{fig:outflow} and \ref{fig:slices}, the effects of individual and groups of clusters can be seen as cleared out, low density holes in the disk. In some cases these holes are the result of ongoing cluster feedback, as seen on the far left of the disk slice in Figure~\ref{fig:slices}. Here, an individual superbubble driven by a high-mass cluster is actively blowing out gas, and is characterized by a hot ($10^6\K$), very high velocity inner free-wind region, surrounded by a region of even hotter shocked gas ($10^8\K$). It is the interaction between gas injected by the cluster (identifiable by its pure yellow color in the fourth panel of Figure~\ref{fig:slices}) and pre-existing disk and halo gas that gives rise to the very high temperatures seen in the simulation.

The fifth panel of Figure~\ref{fig:slices} shows the color variable, a passive scalar field that traces either cluster ejecta (a color of 1) or ISM material (a color of 0). As was also noted in our analysis of the CGOLS IV simulation, very few regions in the outflow appear to consist of pure cluster ejecta or pure ISM material; beyond a radius of a few kpc, the gas in the wind is well mixed. Nevertheless, traces of gas origin in the wind can be seen via the correlations between velocity and color. The less the cluster ejecta has mixed with other gas, the higher its velocity. This is consistent with our results from the CGOLS IV model, in particular Figure 17 \citep{Schneider20}.

\pagebreak

\subsection{Radial Profiles}\label{sec:profiles}

We now turn to a statistical description of the properties of gas in the wind as a function of radial distance from the center of the domain. In keeping with the analytic model presented in \citetalias{Schneider20}, we divide the gas in the wind into phases, where ``hot" contains all gas with $T > 5\times10^5\K$, and ``cool", $T < 2\times 10^4\K$. We then analyze the gas in radial shells of width $\Delta r = 0.1\kpc$, calculating the mean, median, and 25th and 75th quantile of all gas in the shell for each physical quantity of interest. These quantities include the number density,
\begin{equation}
    n = \frac{\rho}{\mu m_\mathrm{p}},
\end{equation}
calculated from the mass density $\rho$ and assuming a mean molecular weight of $\mu = 0.6$, as appropriate for ionized gas; the radial velocity, $v_r$; the thermal pressure, 
\begin{equation} 
    P = (E_\mathrm{tot} - E_\mathrm{kin})(\gamma - 1), 
\end{equation}    
where $E_\mathrm{tot}$ is the total gas energy, $E_\mathrm{kin} = \frac{1}{2}\rho \mathbf{v}^2$ is the gas kinetic energy, and $\gamma$ is the adiabatic index of the gas, assumed to be $\frac{5}{3}$; the temperature, 
\begin{equation}
    T = \frac{P}{n k_B};
\end{equation}
the scalar variable,
\begin{equation}
    s = \rho_s / \rho;
\end{equation}
the sound speed,
\begin{equation}
    c_s = \sqrt{\gamma P / \rho};
\end{equation}
the Mach number, 
\begin{equation}
    \mathcal{M} = v_r / c_s;
\end{equation}
and the entropy, which we calculate as 
\begin{equation}
    K = P n^{-\gamma}.
\end{equation}    

\begin{figure*}
    \centering
    \includegraphics[width=\linewidth]{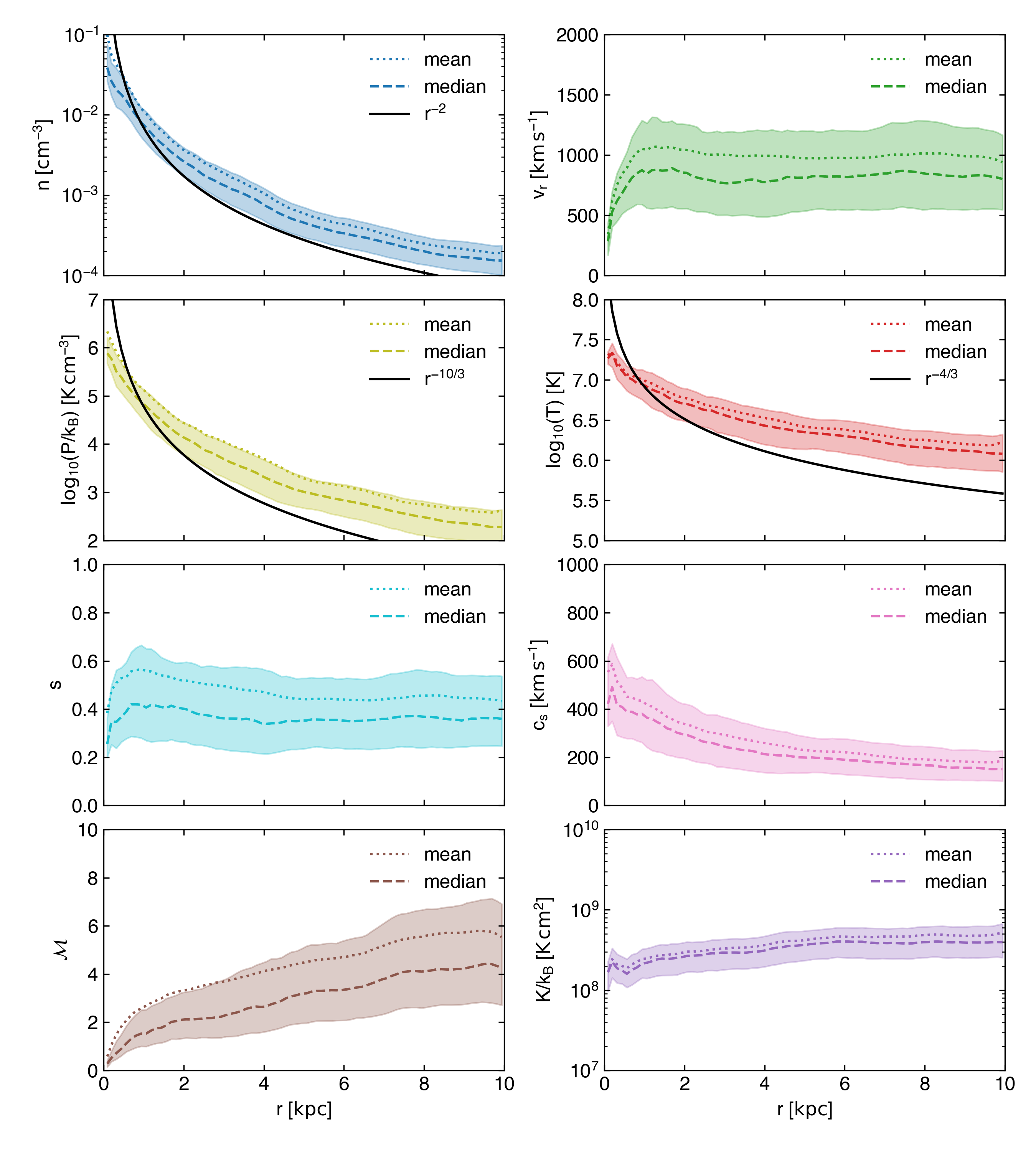}
    \caption{Radial profiles for the hot phase ($T > 5\times10^5\K$) within in a $30^\circ$ half-opening angle cone. From top-left to bottom right, profiles are displayed for number density, radial velocity, pressure, temperature, scalar color, sound speed, mach number, and entropy. Each plot shows the mean, median, and 25th and 75th percentiles for each quantity, and all averages are density-weighted.}
    \label{fig:profiles_hot}
\end{figure*}

\begin{figure*}
    \centering
    \includegraphics[width=\linewidth]{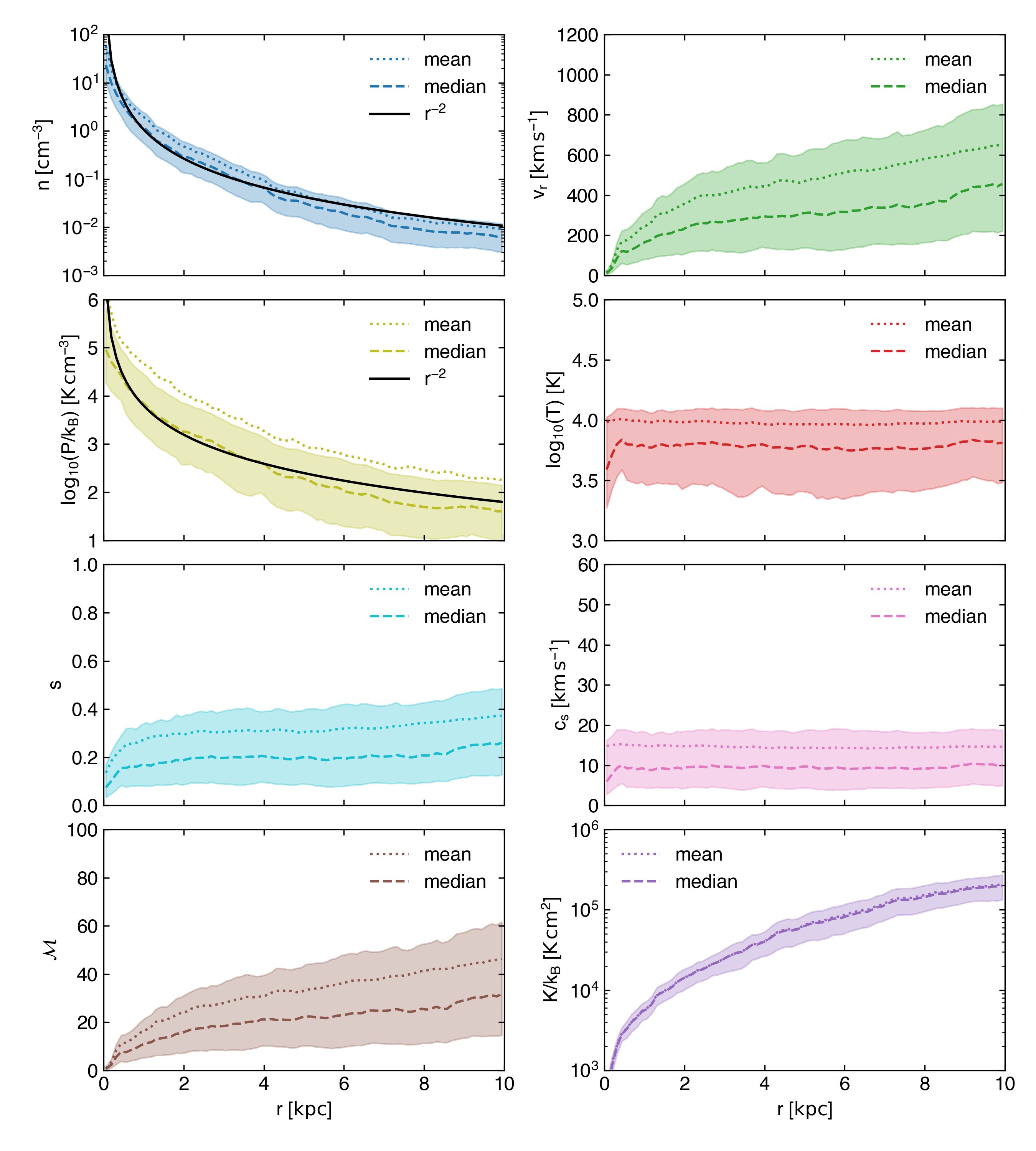}
    \caption{Same as Figure \ref{fig:profiles_hot}, but for the cool phase ($T < 2\times10^4\K$).}
    \label{fig:profiles_cool}
\end{figure*}

For each variable except the number density, we calculate density-weighted statistics in the bin. For example, for the mean radial velocity,
\begin{equation}
    v_{r, \mathrm{av}} = \frac{\Sigma (v_{r, i} n_i)}{n_\mathrm{av}},
\end{equation}
where the sum and the average are taken over all of the cells in the radial shell within a cone of half-opening angle $\Omega = 30^\circ$. Sums are taken both above and below the disk. We perform the analysis within a biconical region for comparison with our previous work. These profiles are plotted for the hot phase in Figure~\ref{fig:profiles_hot}, and for the cool phase in Figure~\ref{fig:profiles_cool}.

As in our central wind model, we find that for the hot phase, the profiles for number density, pressure, and temperature are significantly flatter than those expected for an adiabatically expanding, thermally-driven wind, such as that described by \cite{Chevalier85} (hereafter CC85). Expected slopes for the pure adiabatic expansion model are shown as thick black lines on each of the relevant panels. We attribute these flatter slopes primarily to the transfer of mass from the cool phase to the hot, which has the effect of flattening the profiles, as described by \cite{Nguyen21} and \cite{Fielding22}. However, we note that the mass outflow rates do not show evidence of \textit{net} mass transfer from the cool to the hot phase in this simulation (see Section \ref{sec:fluxes}).

One possible explanation for this apparent contradiction is the effect of local versus global behavior in the hot phase. Figure \ref{fig:slices} shows that there are large regions in the outflow at all radii that are in the cool phase. If hot phase material is cooling out in some regions, it will no longer contribute to the average profiles shown in Figure \ref{fig:profiles_hot}, which are insensitive to the total amount of mass or volumetric area in a given radial bin. Thus, it can be physically consistent that in local regions where the hot phase persists, it shows the effects of mass transfer from the cool phase (as is also evident from the color panel in Figure \ref{fig:slices}), while the global behavior indicates that in aggregate, the mass in the hot phase is not increasing as a function of $r$.

Several other features of the hot phase profiles are of note. First, the density-weighted radial velocities for the hot gas are lower than would be expected from a pure adiabatic expansion model. The expected asymptotic velocity in a CC85-type model is
\begin{equation}
    v_{\infty} = \left(\frac{2\dot{E_\mathrm{inj}}} {\dot{M_\mathrm{inj}}}\right)^{1/2}.
\end{equation} 
As shown in Figure~\ref{fig:S99rates}, the energy and mass injection rates vary over time for each cluster, but using the average values this asymptotic velocity is of order $2600\kms$. Indeed, this is approximately the velocity that is seen within the superbubble on the left side of the disk in Figure~\ref{fig:slices}. However, the interaction of the hot wind with the rest of the outflowing material, in particular the slower-moving cool phase, has the effect of draining kinetic energy from the hot phase, both by conversion to thermal energy in shocks, and by transferring momentum to the cool phase. The net result is a substantially slower-moving hot phase with a relatively constant mean velocity of around $1000\kms$ (though we note there is a large spread at any given radius).

The panel for the scalar value, $s$, gives further insight into this lower velocity. While pure cluster ejecta has a scalar value of 1, the mean scalar value in the hot phase of the outflow is $\sim0.5$, indicating that 60\% of the mass being carried in this phase was originally part of the ISM. Thus, the total momentum originally carried by the hot phase is now being shared with gas that had an initial radial velocity of 0, and the net effect is velocities of order $0.5\times2600\kms \approx 1300\kms$, which is much closer to the median value in the hot phase (see also the discussion in Section 3.2 of \citetalias{Schneider20}). While this estimate is slightly higher than the velocities in Figure~\ref{fig:profiles_hot}, it does not take into account further deceleration due to shocks, nor any additional energy losses due to radiative cooling, which may be large, especially in the mixed phase of gas where the momentum is transferred \citep{Fielding20a}.

Finally, we observe that the sound speed in the hot gas is dropping with radius, leading to a rising Mach number for the wind, as expected for an expanding wind model. However, the entropy is also rising slightly, which is unexpected for an adiabatically-expanding model (though consistent with our previous results). We again attribute this to mass transfer into the hot phase, as the increasing entropy profile is a feature that is reproduced by including a mass source term in the model of \cite{Nguyen21}. However, we note that the rise in the entropy profile is not as significant as was observed in our previous work, and it flattens out at larger radii, indicating less mass-loading of the hot phase than was seen in the CGOLS IV simulation. We return to this point in Section \ref{sec:fluxes}.

\begin{figure}[t!]
    \centering
    \includegraphics[width=\linewidth]{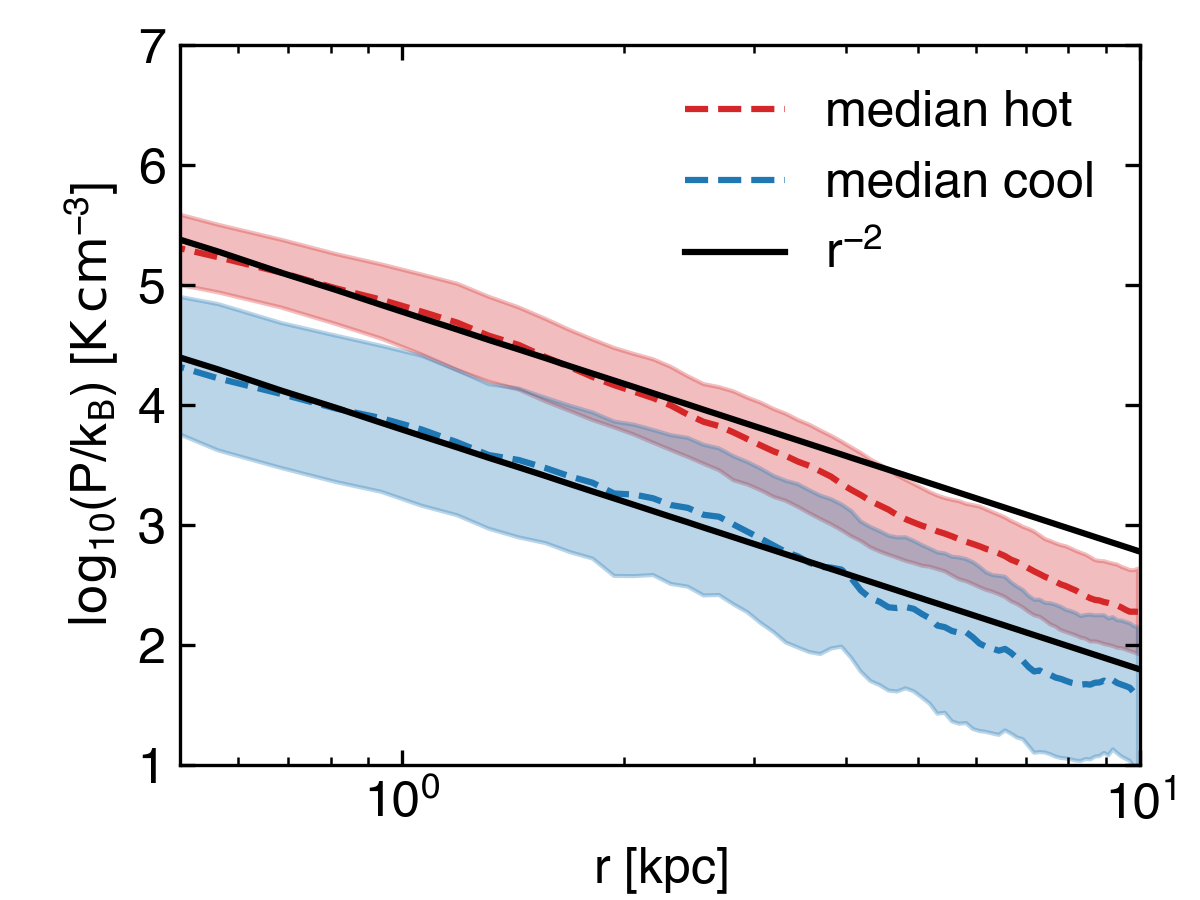}
    \caption{Pressure profiles for the hot and cool phases of the wind at 30 Myr. Black lines show $P\propto r^{-2}$ profiles, normalized at $1\kpc$.}
    \label{fig:p_compare}
\end{figure}

Figure~\ref{fig:profiles_cool} shows the corresponding radial profiles for the cool phase. Naturally, the temperature of the gas in this phase is all near the cut-off for the cooling curve at $T = 10^4 \K$, and thus the density and pressure profiles are very similar. The density profile is fit reasonably well by a simple expansion model, $n \propto r^{-2}$, as shown by the solid black line in the first panel. While this may seem reasonable at first glance, it is not entirely clear that this should be the expected scaling. Because the cool phase is isothermal, we expect the density profile to follow the pressure, which, for cool clouds embedded in a hot medium, one might expect to be in equilibrium with the hot phase. However, the hot phase pressure profile, while not as steep as the $r^{-10/3}$ slope predicted by adiabatic expansion, is still significantly steeper than the $r^{-2}$ slope observed for the cool phase. We show this explicitly in Figure~\ref{fig:p_compare}, which directly compares the hot and cool pressure profiles. Evidently, the two profiles are not directly coupled. In addition, the cool phase has an order of magnitude lower pressure at small radii, though the two phases get closer at larger distances.

We can understand this decoupling if much of the gas in the cool phase is not actually in sonic contact with the hot background wind. This can be the case if clouds have significantly shorter cooling times than their sound crossing times, which is true in our model. For example, for clouds at $R\sim 1\kpc$ with a typical number density $n \approx 0.1 - 1\cc$, the cooling time is short, of order $10^4 - 10^5\yr$. The sound crossing time for clouds at the resolution limit of the simulation, with $R_\mathrm{cl} \approx 2\Delta x \approx 10\pc$, is a factor of 10 larger than this, $t_\mathrm{cross} \approx 1\Myr$. Most clouds are significantly larger than this. Thus, the pressure profile for the cool phase does not have to be in equilibrium with the hot phase, and is free to follow the expected profile for isothermal radial expansion, $r^{-2}$.


Figure \ref{fig:profiles_cool} also shows that the cool phase velocities are increasing with radius on average. Although some of this apparent increase could be a result of lower velocity gas dropping out of the outflow at larger radii, we do not see a large enough spread in cool gas velocity at small radii for that possibility to fully explain the apparent acceleration. In keeping with the model outlined in \citetalias{Schneider20}, we instead attribute this increase primarily to momentum transfer into the cool phase from the hot phase via mixing, which is a function of the distance the cool gas has traveled. As shown in the scalar panel, the cool phase has an average scalar value of 0.3 by 1kpc, which increases to 0.4 by 8 kpc, indicating that mass (and corresponding momentum) has been transferred to it from the high-scalar hot gas (which starts in the clusters with a scalar value of 1), and continues to be transferred as the cool gas moves out. While this may seem to contradict our earlier conclusion that mass is being transferred into the hot phase, we emphasize that these are global averages, while the actual mixing processes responsible for mass transfer from one phase to another are local. Thus, it is possible for individual cool clouds to either gain or lose mass to the hot phase in a way that results in no net mass transfer from one phase to the other, while still allowing the cool gas to gain momentum on average.

The cool phase has a roughly constant temperature, and therefore a roughly constant sound speed of $\sim 10\,\mathrm{km}\,\mathrm{s}^{-1}$. The increase in velocity then results in an increasingly supersonic cool outflow as a function of radius, as seen in the Mach number panel. We also see from the entropy panel that the cool phase is gaining entropy as it moves outward, which is likely due to a combination of lower entropy gas dropping out of the outflow at larger $r$, as well as continued mixing with the higher entropy hot phase.

\subsection{Outflow Rates}\label{sec:fluxes}

\begin{figure}
    \centering
    \includegraphics[width=\linewidth]{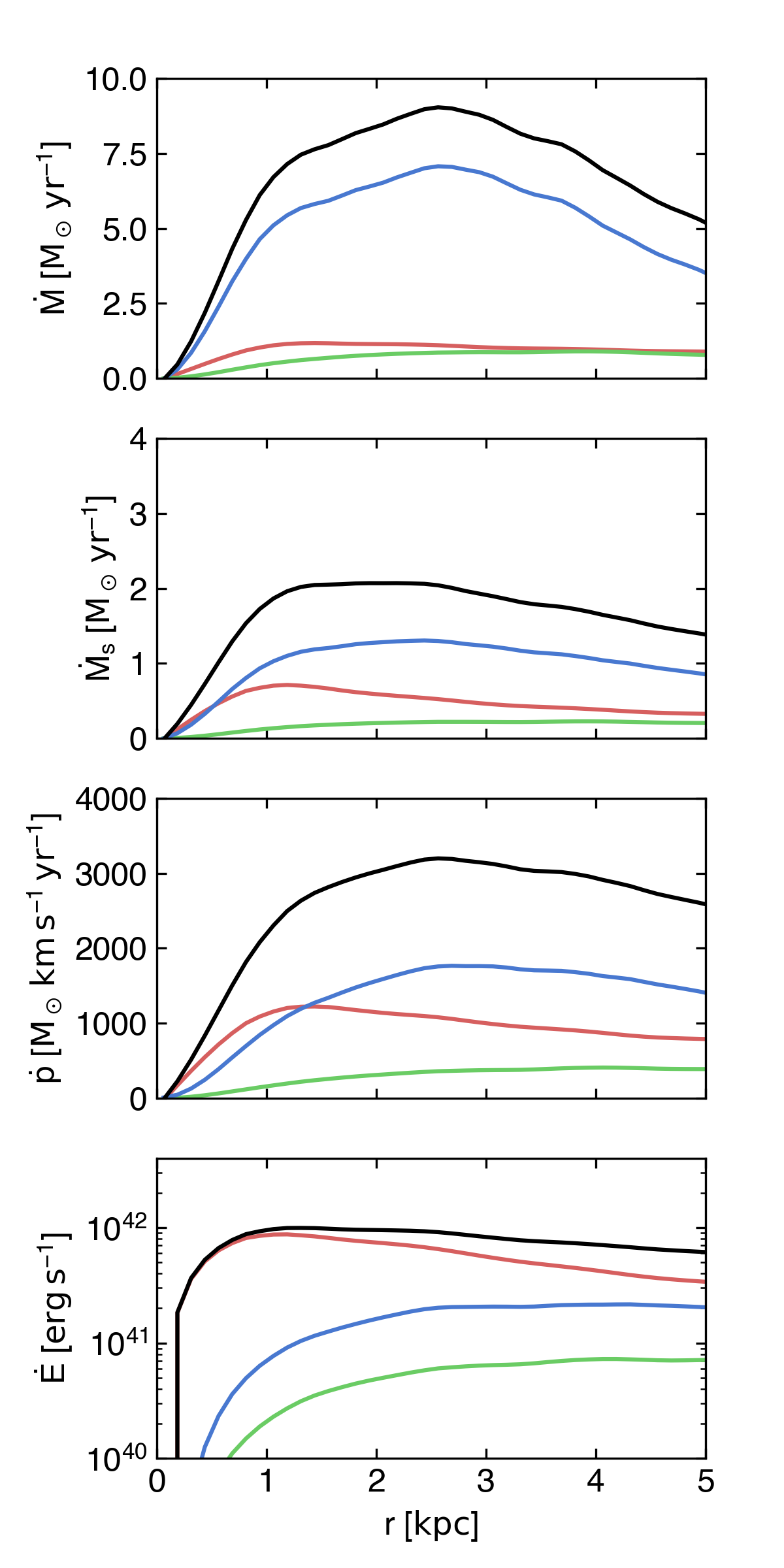}
    \caption{Radial outflow rates of mass, scalar mass, momentum, and energy in spherical shells, excluding positions within $5^\circ$ of the disk midplane. Rates are broken down by phase, with cool gas in blue ($T < 2\times 10^4\,\mathrm{K}$), intermediate gas in green ($2\times 10^4\,\mathrm{K} < T < 5\times 10^5\,\mathrm{K}$), and hot gas in red ($T > 5\times 10^5\,\mathrm{K}$). The total outflow rate in each quantity is shown in black.}
    \label{fig:full_fluxes}
\end{figure}

We now turn our attention to the outflow rates measured in the simulation after 30 Myr of feedback. In Figure~\ref{fig:full_fluxes} we plot the total mass, scalar mass, momentum, and energy outflow rates through radial shells, excluding locations with azimuthal angles within $\phi = 5^\circ$ of the disk midplane. Following the procedure outlined in \citetalias{Schneider20}, rates are calculated by integrating the fluxes within shells with bin width $\Delta r = 0.125\,\mathrm{kpc}$, and smoothed over 3 radial bins. No time averaging is applied.

In Figure~\ref{fig:full_fluxes} we calculate fluxes over (almost) an entire sphere rather than only in a $60^\circ$ biconical region in order to get a better sense of the total outflow rates and how they compare to commonly measured loading factors. We define the mass loading factor, $\eta_m$, as the measured mass outflow rate through a shell relative to the star formation rate, 
\begin{equation}
    \dot{M}_\mathrm{outflow} = \eta_m \dot{M}_\mathrm{SFR}.
\end{equation}
We see from Figure~\ref{fig:full_fluxes} that the radially-averaged mass outflow rate peaks at approximately 2.5 kpc, at a rate close to $10\Msun\,\mathrm{yr}^{-1}$. With an assumed star formation rate of $20\Msun\,\mathrm{yr}^{-1}$, this corresponds to a peak mass loading rate of $\eta_m = 0.5$. However, at larger radii the outflow rate drops off, falling to $\eta_m = 0.25$ by a distance of 5 kpc. This is consistent with the presence of a low-$z$ fountain flow, in which low velocity material that is outflowing at small radii begins to fall back toward the disk at larger radii. Similar trends are also observed in simulations with planar geometry \citep[e.g.][Fig. 6]{Kim20a}. The fact that this feature is only seen in the cool phase is also physically consistent with a fountain model, given that the hot phase is traveling at velocities above the escape speed and should not fall back.

Scalar mass-loading tells a similar story. Because the passive scalar is injected only in clusters, we can calculate its loading factor relative to the hot gas injection -- this serves as a proxy for metallicity, since the metal enriched gas ejected by supernovae is expected to start in the hot phase. The cluster injection rate varies as a function of time, but is $0.175\,\dot{M}_\mathrm{SFR}$ when averaging across all clusters at all ages. Thus, the injected scalar mass is approximately $3.5\Msun\,\mathrm{yr}^{-1}$. At $r < 2.5\kpc$ the observed total scalar outflow rate reaches $2\,\Msun\,\mathrm{yr}^{-1}$, approximately 60\% of the total, and drops to 40\% at 5 kpc. The implication is thus that the metal loading in the outflow is higher than the total mass loading -- a reasonable result if the hot phase carries preferentially more metals -- but that at least 60\% of the newly-generated metals will remain in the ISM, in large part due to efficient mixing of the ejecta with the cool phase at small radii. This result is consistent with the cool gas profiles in Figure \ref{fig:profiles_cool}, which show that roughly 10 - 50\% of the cool gas is composed of scalar material (i.e. was once injected cluster mass). Thus, as cool gas drops out of the outflow at larger $r$, the scalar mass loading decreases.

The third panel of Figure \ref{fig:full_fluxes} shows the momentum outflow rates in the three phases. We can define a ``momentum loading" factor, $\eta_p$, by comparing the total outflowing momentum rate to a reference based on the terminal momentum injected into the ISM from a single supernova, $p_\mathrm{ref} = 1.25\times10^5\Msun\kms$ \citep[e.g.][]{Kim20b, Pandya21}. Then the measured momentum in the outflow can be related to the star formation rate by assuming there is one supernova for every $100\Msun$ of star formation,
\begin{equation}
        \dot{p}_\mathrm{outflow} = 1.25\times 10^{3}\Msun\kms\mathrm{yr}^{-1}\, \eta_p \left[\frac{\dot{M}_\mathrm{SFR}}{\Msun\,\mathrm{yr}^{-1}}\right].
\end{equation}
Given this reference rate, we see that the total momentum loading factor in the outflow is low, approximately $\eta_p = 0.1$. At small radii, more of the momentum is carried by the hot phase, but at large radii, enough momentum has been transferred that the cool phase dominates the momentum outflow rate.

Finally, the fourth panel displays the energy outflow rates. Similarly to the momentum loading, we define the energy loading factor, $\eta_E$, relative to the star formation rate and assuming that every $100\,M_\odot$ of  star formation produces one $10^{51}$ erg supernova\footnote{We note that this definition of energy loading was chosen to allow more straightfoward comparison of the measured outflow properties with other work, but its formulation is slightly inconsistent with the actual $\dot{E}_\mathrm{inj}$ values used in our clusters, which is on average $\dot{E}_\mathrm{inj} = 3.9\times 10^{41}\,\mathrm{erg}\,\mathrm{s}^{-1}$. Thus, the average \textit{injected} energy loading factor is $\eta_E = 1.25$, rather than the typical $\alpha = 1$.}, giving the relationship
\begin{equation}
    \dot{E}_\mathrm{outflow} = 3.1\times 10^{41}\,\mathrm{erg}\,\mathrm{s}^{-1}\, \eta_E \left[\frac{\dot{M}_\mathrm{SFR}}{\Msun\,\mathrm{yr}^{-1}}\right].
\end{equation}
Comparing this to Figure~\ref{fig:full_fluxes}, we see that the total energy loading is also low, approximately $\eta_E = 0.1$ at 5 kpc. Although the hot phase does carry more energy out than the cool, the hot phase energy loading decreases as a function of radius, while the cool phase increases and then steadies around 3 kpc. Several conclusions can be drawn from these trends. First, the cool phase that persists in the outflow must continue to gain energy as it moves out, since the overall mass outflow rate is dropping as a function of $r$. This is corraborated by the steadily increasing velocity and scalar fractions seen in Figure ~\ref{fig:profiles_cool}. Second, given the relatively steady hot phase mass outflow rate and velocity, the dropping hot phase energy loading may indicate continued losses due to radiative cooling in the outflow, not just near the base. Finally, the similar rise in intermediate phase mass and energy outflow rates as a function of $r$ indicates that ongoing mixing is playing a role in both the cool gas acceleration and the hot gas energy loss. These trends differ from our previous work, in which the hot phase energy outflow rates remained flat as a function of distance, and the cool phase energy loading decreased as cool gas was destroyed.

\subsection{Comparison to the CGOLS IV model}\label{sec:comparison}

Given the differences in many of these trends with our previous work, in particular to the central burst model (CGOLS IV), we now turn to a more detailed comparison between the two simulations. Because the CGOLS~V model is qualitatively similar to the CGOLS IV model described in \citetalias{Schneider20}, in this Section we focus on the differences.

As mentioned in Section~\ref{sec:simulations}, the only difference in setup between the two simulations are the details and spatial distribution of the cluster feedback. CGOLS~IV employed a more centrally-concentrated cluster distribution modeled after a nuclear starburst, with all clusters placed within the central $R = 1\kpc$. In CGOLS V, the radial cluster distribution follows the exponential surface density distribution of the disk gas, with a scale radius of 1 kpc. Averaged over the whole disk, this results in a star formation rate surface density approximately 20 times higher for CGOLS IV, although the ratio is lower in the central regions and formally zero at radii greater than 1 kpc. In addition, the clusters in CGOLS IV were all the same mass, $10^7\Msun$, and were on for only $10\Myr$, whereas the clusters in the present simulation have a power law distribution of masses between $10^4\Msun$ and $5\times10^6\Msun$, and are on for 40 Myr. Therefore, although there are many more clusters in the CGOLS V simulation, individual clusters are much less powerful than those used in CGOLS IV.

\begin{figure*}
\centering
\includegraphics[width=0.45\linewidth]{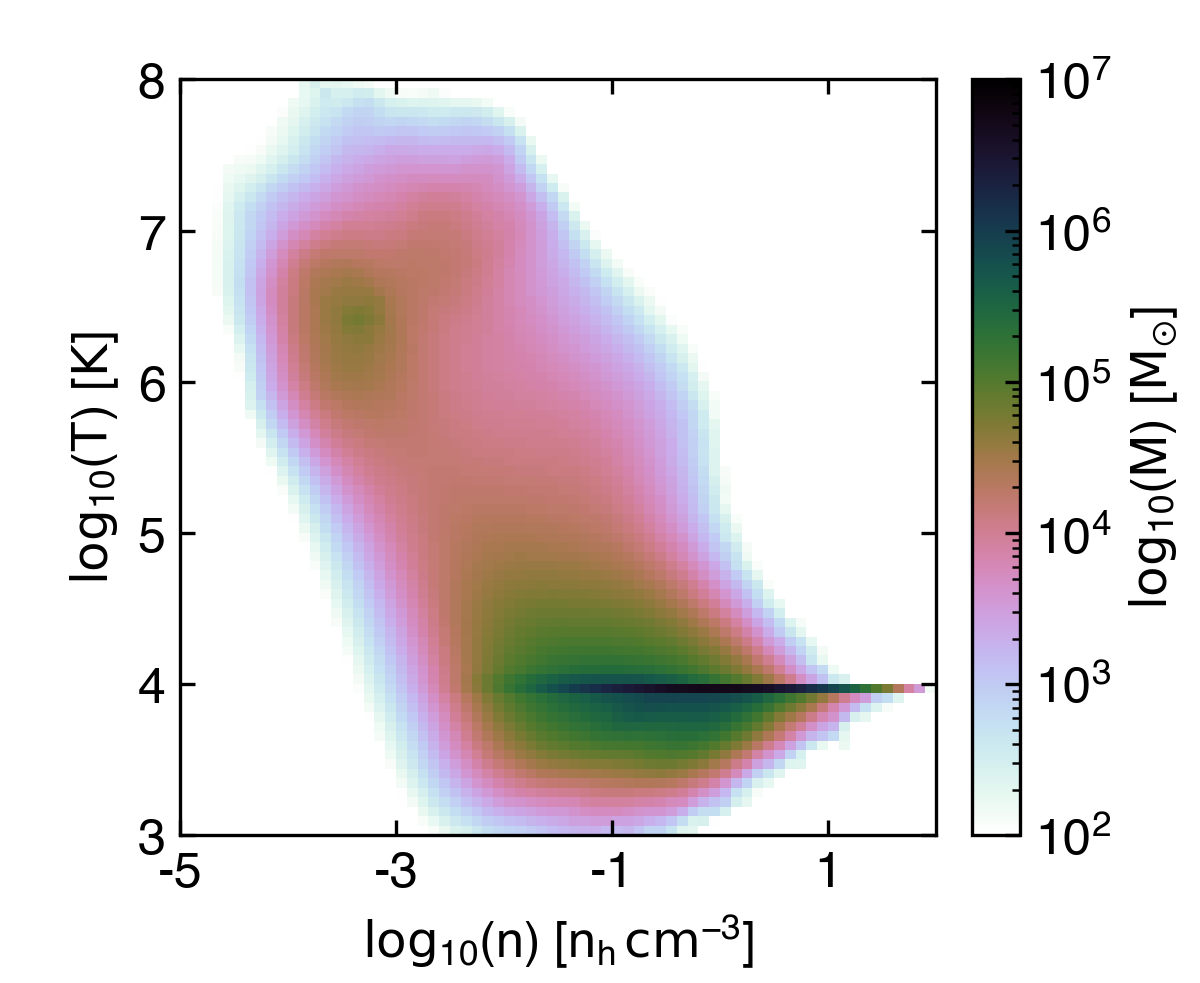}
\includegraphics[width=0.45\linewidth]{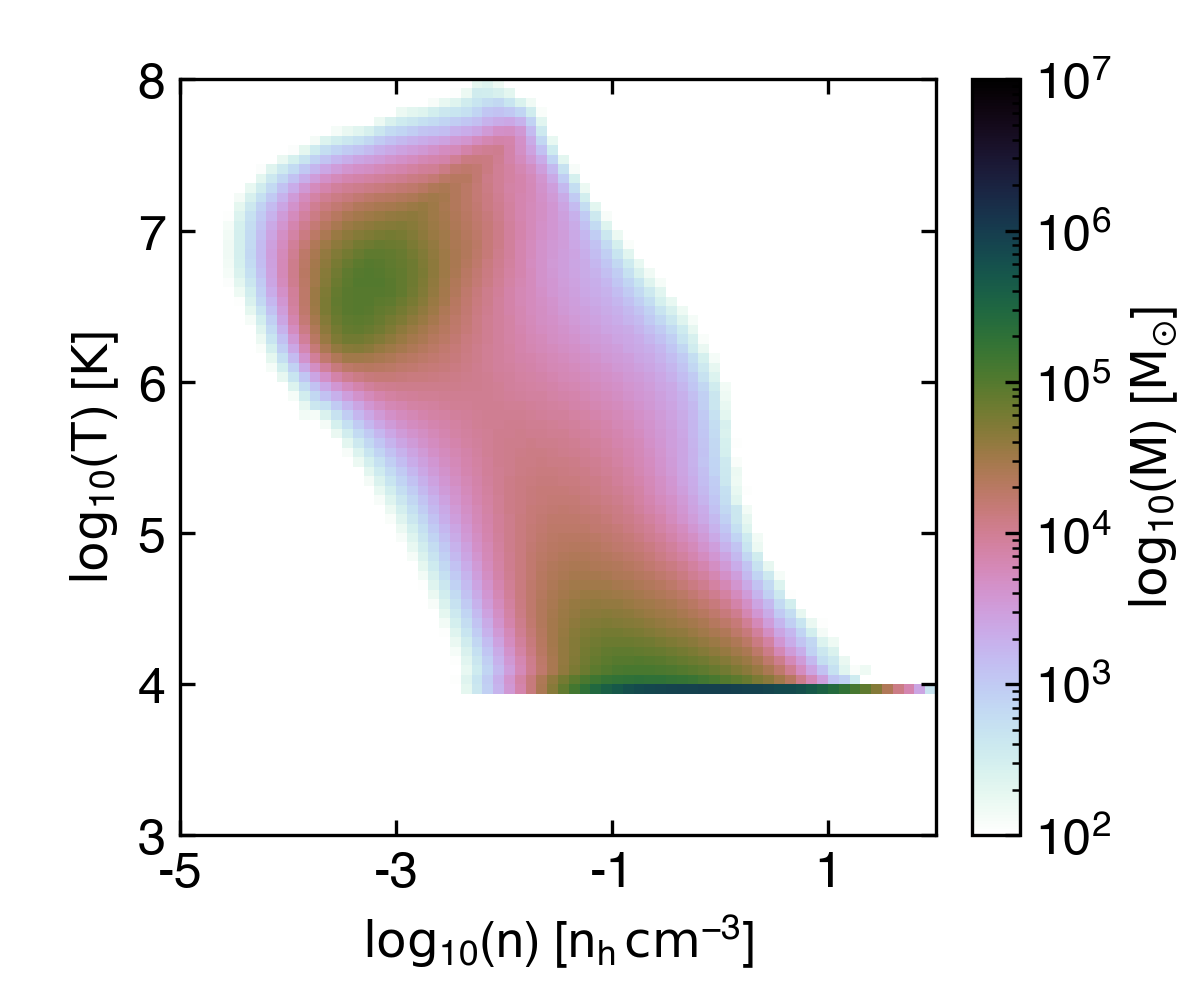}
\caption{Density-temperature phase plots for the distributed cluster simulation CGOLS V (left) and the central burst simulation CGOLS IV (right). Bins are weighted by mass. All the gas in the simulation volume is included except cells with azimuthal angles within $5^\circ$ of the disk, and heights within $z = 0.5\kpc$ of the disk. Both plots correspond to a snapshot 30 Myr after the start of feedback.}
\label{fig:phase_plots}
\end{figure*}

\begin{figure}
    \centering
    \includegraphics[width=\linewidth]{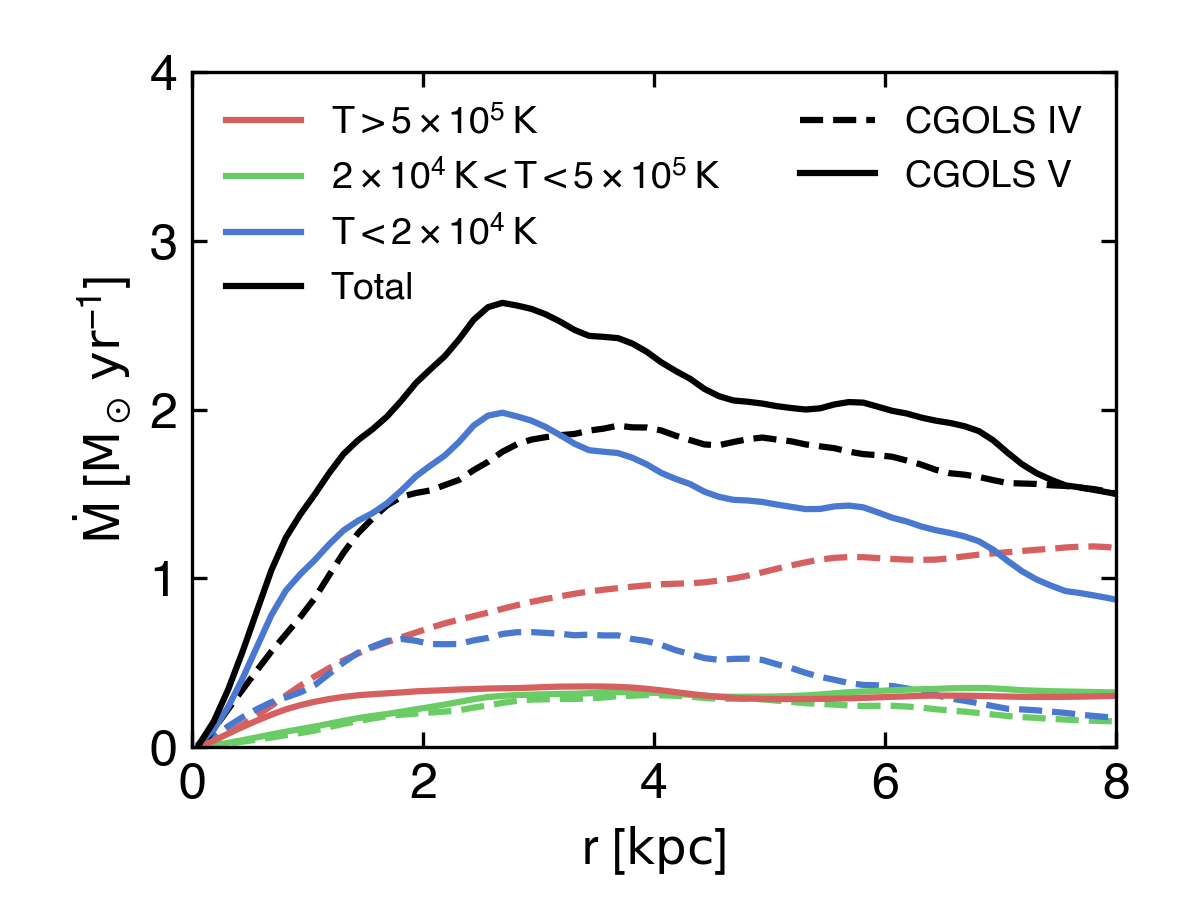}
    \includegraphics[width=\linewidth]{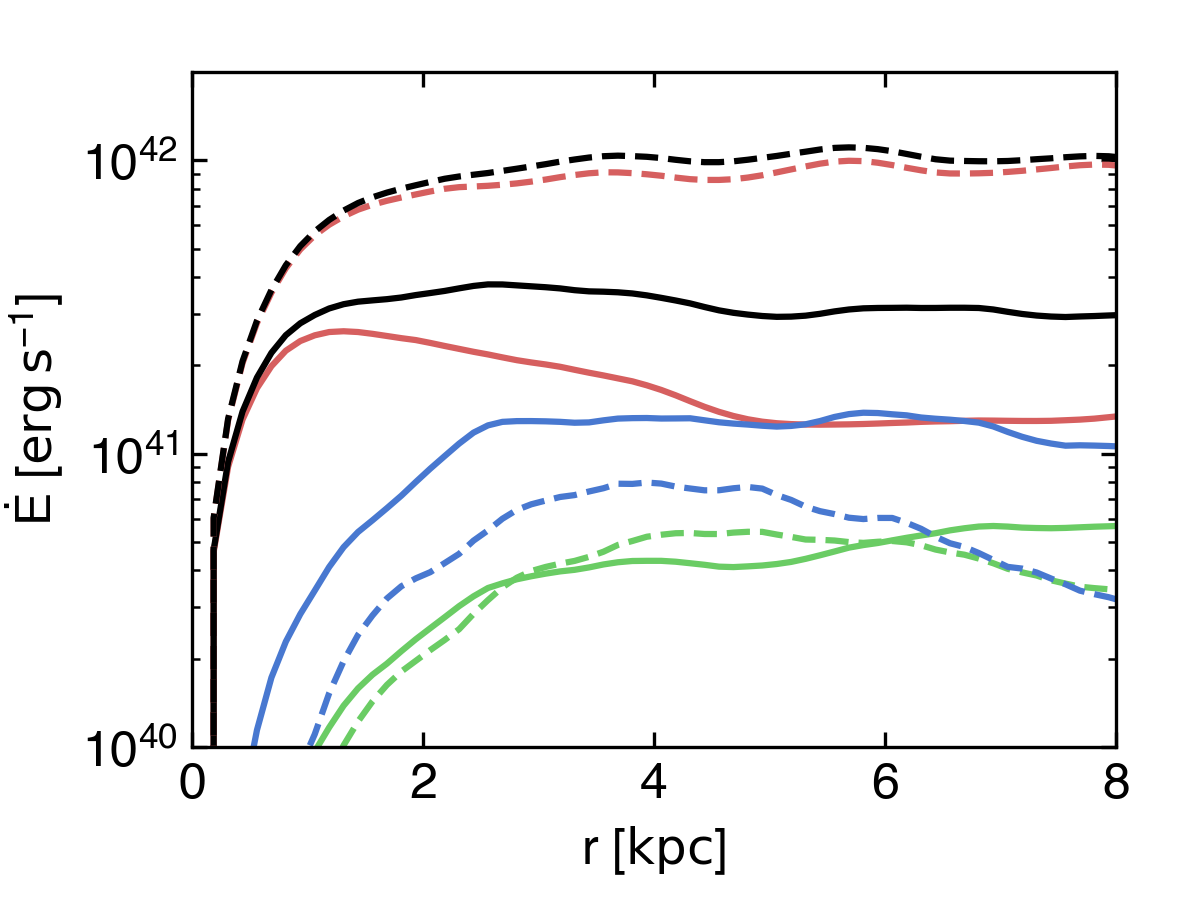}
    \caption{The radial mass and energy fluxes in different phases in the outflow, 30 Myr after cluster feedback began. Fluxes are calculated within a $30^\circ$ half-opening angle cone, for ease of comparison with previous CGOLS models. Solid lines show the fluxes from the distributed cluster feedback simulation (CGOLS V), while dashed lines show the fluxes from the central burst model (CGOLS IV).}
    \label{fig:fluxes_compared}
\end{figure}

\subsubsection{Phase diagrams}

What effect does this have on the properties of the outflow? While the qualitative picture of cool clouds embedded in a hot background flow looks similar, the quantitative picture of the mass in different phases looks somewhat different. We highlight some of these differences in Figures~\ref{fig:phase_plots} and \ref{fig:fluxes_compared}, which compare mass-weighted phase diagrams of the two simulations, and mass and energy fluxes in the two simulations, respectively. As can be seen in Figure~\ref{fig:phase_plots}, a primary difference between the two simulations is the amount of material in the cool phase. While both simulations produce a distinct two-phase outflow with a similar amount of mass in the hot phase, there is far more mass in the cool phase \textit{relative to the hot phase} in the distributed cluster simulation, especially at large azimuthal angles.

In particular, in the distributed cluster model, we find that there is $2.8\times 10^7\,\Msun$ of cool gas within the $60^\circ$ biconical selection region, and $3.2\times 10^6\,\Msun$ in the hot phase -- almost a factor of 10 more mass in the cool phase. These numbers are $7.3\times 10^6\,\Msun$ for the cool phase and $7.0\times 10^6\,\Msun$ for the hot phase in the central burst model. Evidently, while both the distributed starburst and the nuclear starburst are able to produce winds with significant quantities of cool gas, the centrally concentrated model results in a more substantial hot outflow (by a factor of two) and a less substantial cool outflow (by a factor of 4). Expanding the bicone to $170^\circ$, we find that there is $1.5\times 10^8\,\Msun$ of cool gas and $1.0\times 10^7\,\Msun$ of hot gas in the distributed model, versus $2.5\times 10^7\,\Msun$ of cool gas and $1.7\times 10^7\,\Msun$ of hot gas in the central burst model. Again, we see that although there is a similar amount of hot gas between the two simulations (within a factor of 2), there is a factor of 10 more cool gas than hot gas in CGOLS V -- i.e. the primary difference is that the distributed burst is much more efficient at launching cool gas into the outflow.

There is also obviously a large difference in gas below $T = 10^4\K$ visible between the two simulations, but this is not a result of the feedback model, and rather is because CGOLS IV was run with a temperature floor\footnote{Both models were run with the same cooling curve, which cuts off at $10^4\K$. However, the lack of a temperature floor in the CGOLS V simulation means that gas can still expand adiabatically and reach lower temperatures.}. We have rerun a version of the CGOLS V model with the same temperature floor that was used in CGOLS~IV, and confirmed that the total amount of cool gas in both simulations is within a factor of two.

\subsubsection{Outflow Rates}\label{sec:fluxes_compare}

Figure~\ref{fig:fluxes_compared} shows the radial mass and energy fluxes within a $60^\circ$ bicone for both simulations, split into three phases: cool $T < 2\times10^4\K$, intermediate $2\times10^4\K < T < 5\times 10^5\K$, and hot $T > 5\times10^5\K$. Solid lines show fluxes from CGOLS V, and dashed lines show fluxes from CGOLS IV (cc Figure 8 from \citetalias{Schneider20}). Although the two simulations have a remarkably similar total mass outflow rate at $r = 8\,\kpc$ (the largest radial bin we can measure in our conical selection region), in the distributed cluster model a much higher fraction of the outflow is in the cool phase, and thus, the mass flux in that phase is higher relative to the central model. In the distributed model, the mass flux in the cool phase dominates the total mass flux at all radii, and the hot phase mass flux is significantly smaller and steady with $r$, suggesting that there is no net mass transfer from the cool to the hot phase (unlike in CGOLS IV). Interestingly, intermediate temperature mass fluxes between the two models are quite similar. 

Perhaps even more striking are the differences in the energy fluxes. While the CGOLS IV simulation had an energy flux that was dominated by the hot phase at all radii, the CGOLS V model shows a hot phase energy flux that declines with radius, and by the time the outflow reaches $\sim 5\kpc$, the energy flux in the hot and cool phase is approximately equal. There is also a factor of 3 less total energy escaping in the CGOLS V model, despite the fact that the energy injection rates overall are slightly higher. This indicates more substantial losses due to radiative cooling for the distributed cluster model.

We can also get a sense of the degree to which the outflow in CGOLS V is centrally collimated by comparing the fluxes in Figure \ref{fig:fluxes_compared}, which uses a bicone with a  $60^\circ$ opening angle, to those in Figure \ref{fig:full_fluxes}, which uses a $170^\circ$ opening angle. The total mass outflow rate in the smaller cone is approximately $2\Msun\yr^{-1}$ at 5 kpc, versus $5\Msun\yr^{-1}$ for the larger cone. This indicates that the outflow is still quite centrally concentrated, since the ratio of solid angle between the two selections is a factor of 8. The energy outflow rates show a similar ratio, of $3\times 10^{41}\,\mathrm{erg}\,\mathrm{s}^{-1}$ versus $6\times 10^{41}\,\mathrm{erg}\,\mathrm{s}^{-1}$, respectively.

\subsubsection{On the spherical outflow approximation}

\begin{figure}
    \includegraphics[width=\linewidth]{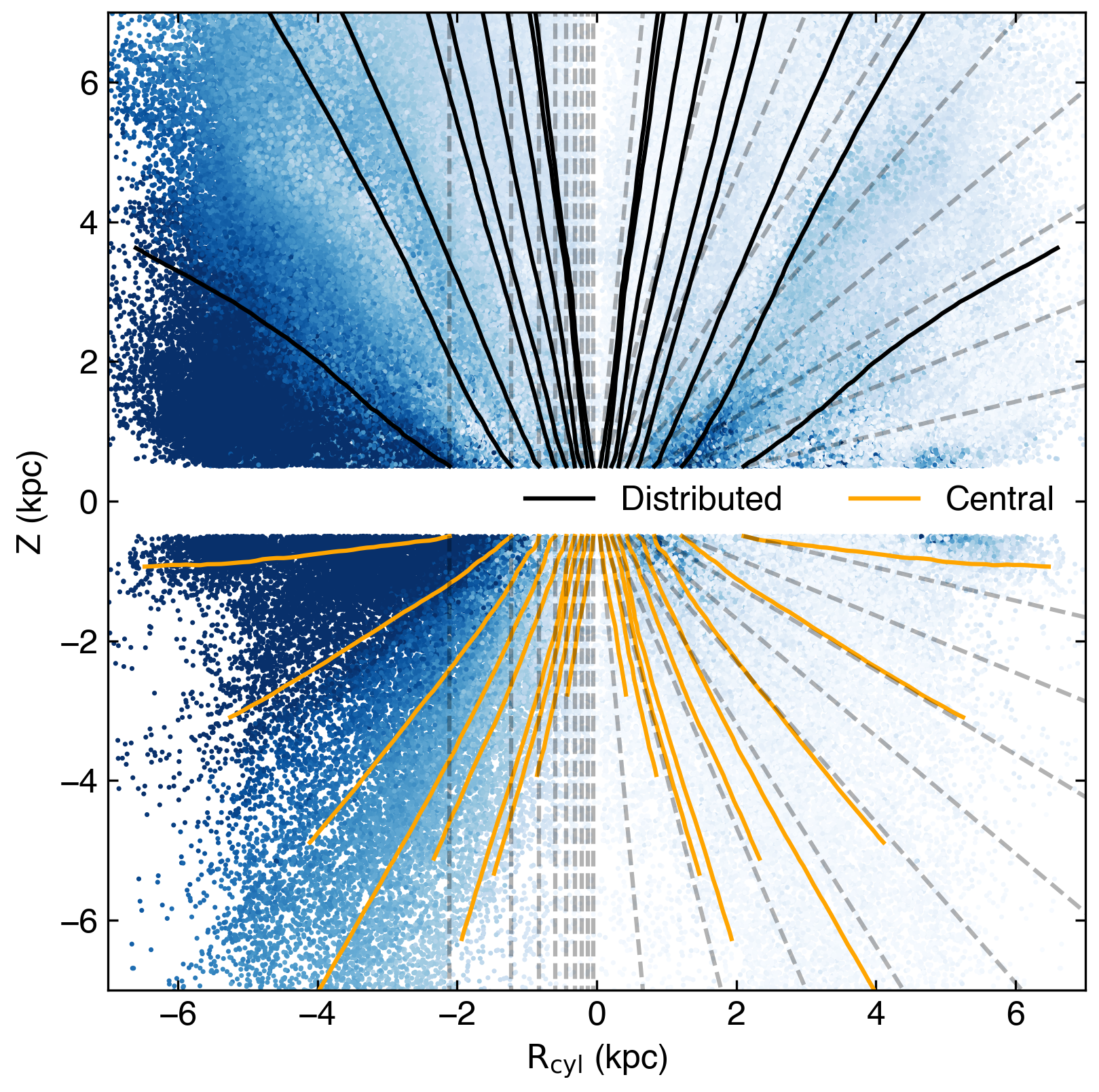}
    \caption{A comparison of cool gas streamlines for the distributed and central models to purely vertical (left) and radial (right) outflow models. Dark solid lines show average streamlines calculated from the velocities of cool clouds within the outflow, identified for each simulation after 30 Myr of feedback. Colored points show the $(R_\mathrm{cyl}, z)$ locations of individual clouds, averaged over the top and bottom of each simulation domain. Density of points indicates the total number of clouds in each simulation, while the color saturation indicates the angular deviation of each cloud's velocity from vertical (left) or radial (right).}
    \label{fig:streamlines}
\end{figure}

As in \citetalias{Schneider20}, we have thus far carried out much of the analysis in this paper in a spherical outflow framework. The primary rationale was to be able to directly compare results, such as profiles and fluxes, to the central burst model. However, the less-centralized placement of clusters in the distributed feedback model warrants some investigation into how good an approximation a radial outflow actually is for each model, as it is conceivable, particularly in the central regions, that the more distributed case could result in streamlines that were closer to vertical. We address this question in Figure~\ref{fig:streamlines}, which plots streamlines for both the CGOLS IV and CGOLS V model.

Figure~\ref{fig:streamlines} uses a catalogue of clouds from each simulation to calculate average streamlines and angular deviations of individual cloud velocities from purely vertical (left) or purely radial (right). ``Clouds" are individual clumps of $T < 2\times 10^4\,\mathrm{K}$ gas that are connected within the simulation volume. The velocity of each cloud is mass-weighted, and only clouds with $\mathrm{abs}(z)$ locations greater than 0.5 kpc are included, to reduce confusion with the disk. The top half of the figure shows data from CGOLS V, while the bottom half shows CGOLS IV. Background grey dashed lines show equivalent vertical and radial streamlines starting in the same location. The density of points on each panel represents the total number of clouds, while the color represents the angular deviation from a vertical (left) or radial (right) outflow.

The larger intensity of color on the left versus the right side of Figure~\ref{fig:streamlines} immediately indicates that a radial outflow is a better fit to the velocities of the cool gas for both simulations. Indeed, the streamlines of the central burst model on the lower right follow the radial streamlines quite closely, and there is very little color saturation. For the distributed model, there is some deviation from radial outflow, particularly at lower $z$ values around $R_\mathrm{cyl} = 1 - 2\,\mathrm{kpc}$, where the radially-averaged density of clusters peaks. However, the streamlines are still much better fit by a radial outflow model, particularly at the outer cylindrical radii and at larger vertical distances from the disk. The largest deviations from vertical can be seen for those regions in the upper-left quadrant of the plot. In particular, clouds at low $z$ in this region may even be part of a fountain of material being ejected at smaller $R_\mathrm{cyl}$ and falling back onto the disk at larger $R_\mathrm{cyl}$, making a vertical outflow a particularly bad approximation.

This figure also demonstrates another important difference between the two outflow models -- there are far more clouds in general in the distributed outflow simulation than in the central burst model, especially in regions at large opening angles. While this result is not surprising, since the central burst model has no clusters at large radii to drive out disk gas, it may begin to account for the differences in cool gas mass in e.g. Figures~\ref{fig:full_fluxes} and \ref{fig:phase_plots}. We explore this idea further in Section \ref{sec:observables}, which addresses how these two models may appear in observational data.

\subsubsection{Origination of differences}

Throughout this Section, we have primarily attributed differences between the two models to the spatial distribution of clusters, by which we mean both the larger physical distribution and much larger total number of clusters in CGOLS V. Both of these factors lead to a much larger surface area of interaction with disk gas for the distributed burst model, which results in many more cool clouds in general, and in particular far more cool clouds at larger radii, as shown in Figure \ref{fig:streamlines}. However, it is worth revisiting the fact that the cluster feedback itself is also significantly different between the two models. In particular, CGOLS IV had clusters that are more massive than the most massive clusters in CGOLS V ($10^7$ versus $5\times10^{6}\,\Msun$, respectively), and they were shorter lived (10 Myr versus 40 Myr, respectively).

We do not expect the cluster lifetimes to have a significant effect, since both models inject the majority of their mass and energy within the first 10 Myr, the average $\dot{M}_\mathrm{cluster}$ and $\dot{E}_\mathrm{cluster}$ are similar, and both simulations are run to a point when the outflow properties are not fluctuating significantly from one snapshot to another. The cluster masses, on the other hand, may play an important role, particularly concerning the interpretation of the energy loading factor in Section~\ref{sec:fluxes_compare}. \cite{Fielding18} demonstrated that for a given gas surface density, larger cluster masses lead to larger average values of $\eta_E$, since large clusters break out of the disk more quickly and are then able to vent their hot gas efficiently into the CGM. In \citetalias{Schneider20}, we argued that this means that the feedback model in CGOLS IV represents a maximally efficient case for energy loading -- all of the star formation is in clusters that have the ability to break out of the disk on a very short timescale. Adding significant additional mass in smaller clusters will then have the effect of ``renormalizing" the measured value of $\eta_E$ downward, consistent with the behavior that is seen in Figure~\ref{fig:fluxes_compared}. This interpretation is also consistent with the fact that the physical characteristics of the hot phase in Figure~\ref{fig:phase_plots} between the two models are quite similar.

Disentangling the effects of cluster distribution versus cluster mass function on the mass loading is more challenging, especially because the measured value of $\eta_m$ in CGOLS V is much more dependent on the distance. We will analyze this degeneracy more thoroughly via additional simulations in future work.

\textbf{}

\begin{figure*}[ht]
    \centering
    \includegraphics[width=0.48\linewidth]{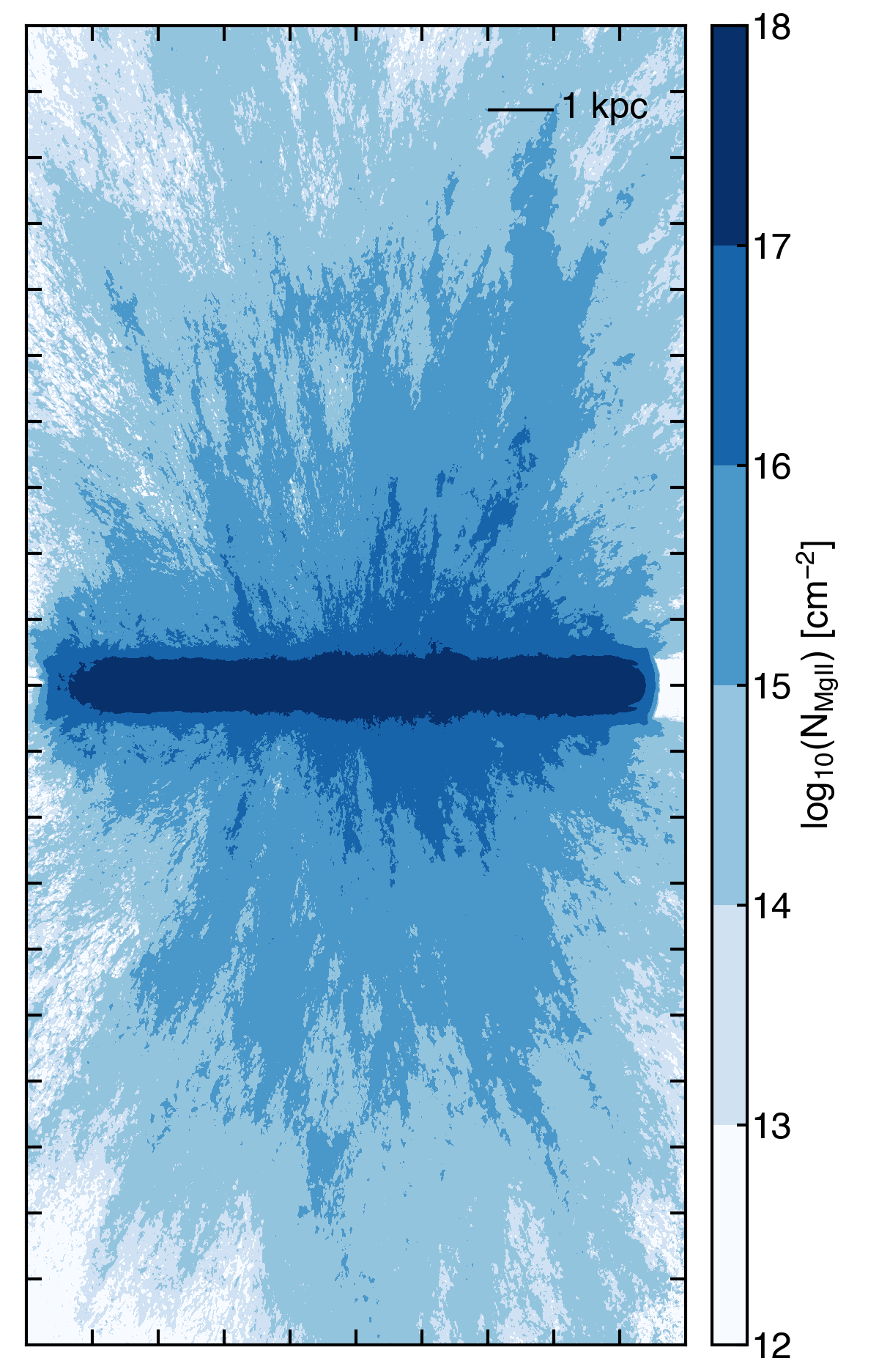}
    \includegraphics[width=0.48\linewidth]{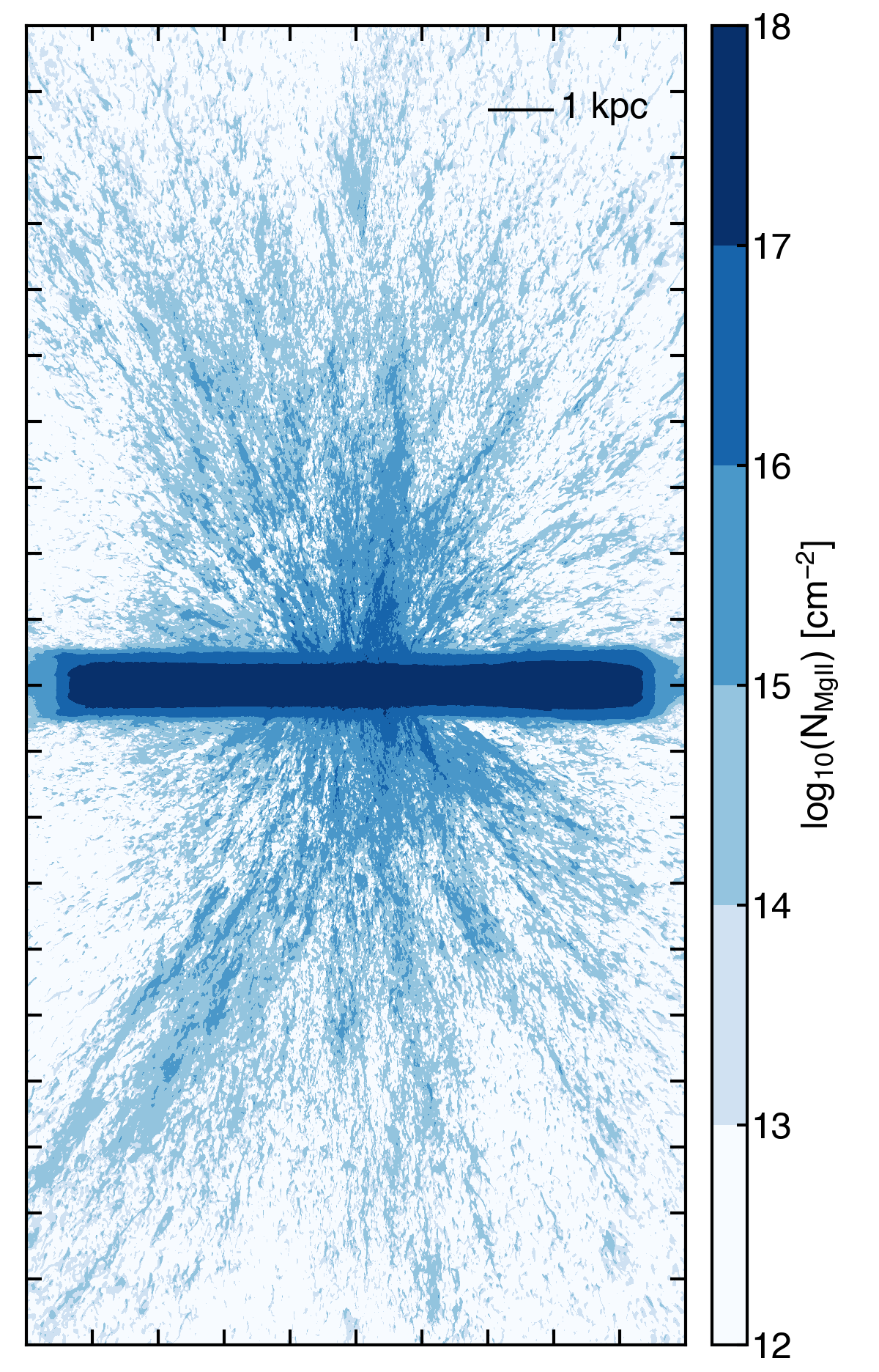}
    \caption{Column density maps of MgII for the distributed feedback model (CGOLS V, left) and the central burst model (CGOLS IV, right) after 30 Myr of supernova feedback.}
    \label{fig:mgII_maps}
\end{figure*}

\section{Observational implications of the CGOLS models}\label{sec:observables}

In this Section we explore several different mock observable properties of both the CGOLS~IV and CGOLS~V simulations. These include column density maps of cool gas, as well as simple mock absorption line spectra for several different ions.

\subsection{Column density maps}

We begin with mock column density maps of cool gas in the simulation, which in this case are just density projections of the simulation data in particular temperature bins. Figure~\ref{fig:mgII_maps} shows $x-z$ projections of column density in Mg II for the distributed model on the left, and the central burst model on the right. The maps were made by assuming solar abundances and solar metallicity for all gas in order to calculate the Mg number density, and additionally assuming that in the cool gas ($T < 2\times 10^4\,\mathrm{K}$) all of the Mg is singly ionized.

The most obvious difference between the two simulations is the extent and covering fraction of high column density cool material. Although both simulations had the same ``star formation rate" of $20\,\Msun\,\mathrm{yr}^{-1}$, the distributed cluster model produces more cool gas at larger radii, indicating that more of the cool clouds are making it farther from the disk in this model. This is consistent with an interpretation in which the rising hot gas mass flux for the central feedback simulation shown in Figure~\ref{fig:fluxes_compared} is a result of mass being transferred from the cool phase to the hot. Thus, at large radii in the central model, the mass in cool clouds is significantly depleted. No analogous rise is seen for the hot gas outflow rate for the distributed cluster simulation in either Figure~\ref{fig:full_fluxes} or Figure~\ref{fig:fluxes_compared}, suggesting that either there is no mass transfer to the hot gas (which is slightly inconsistent with our interpretation of the radial profiles in Figure~\ref{fig:profiles_hot}), or there is comparable depletion of the hot phase onto clouds.

\begin{figure*}[ht!]
    \centering
    \includegraphics[width=0.45\linewidth]{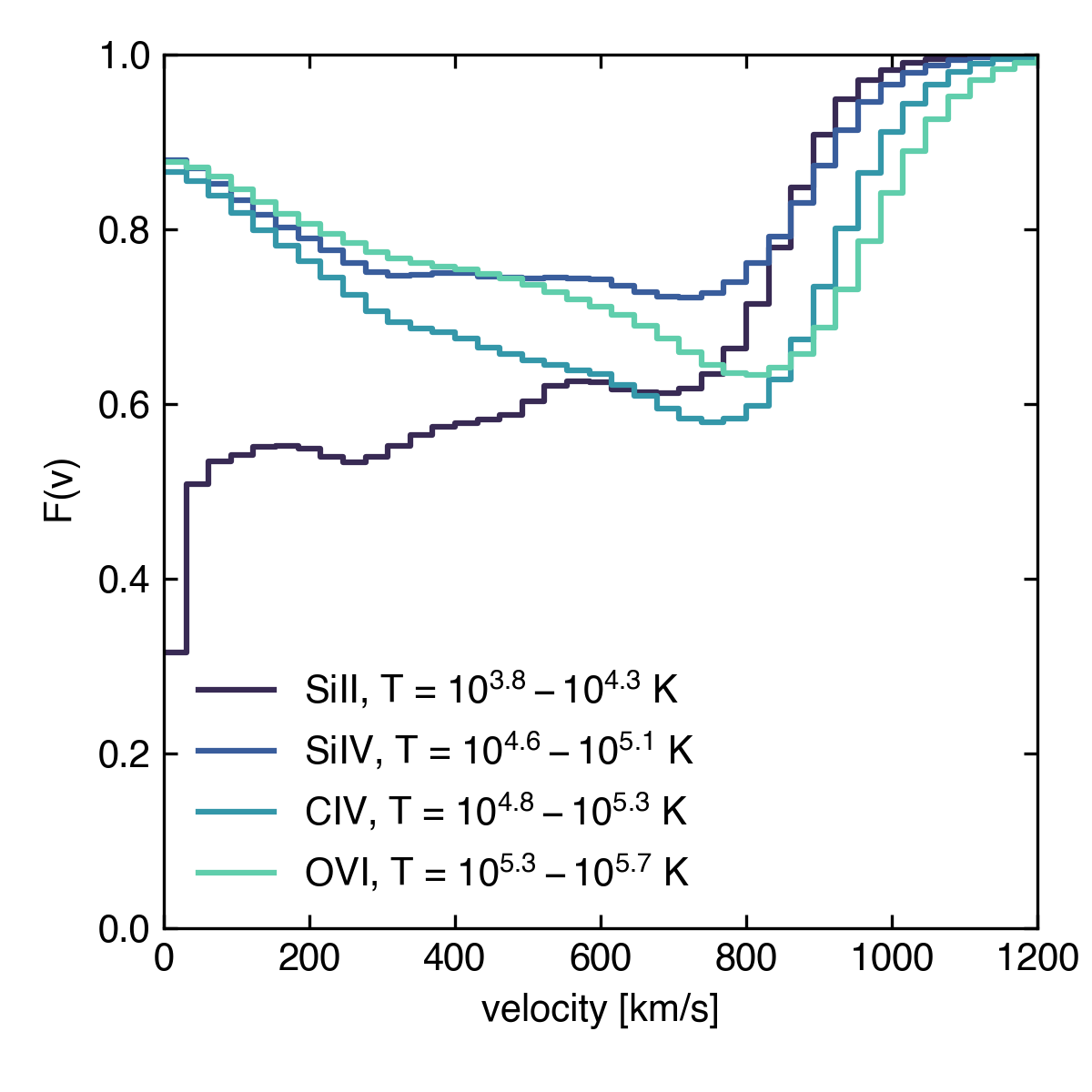}    
    \includegraphics[width=0.45\linewidth]{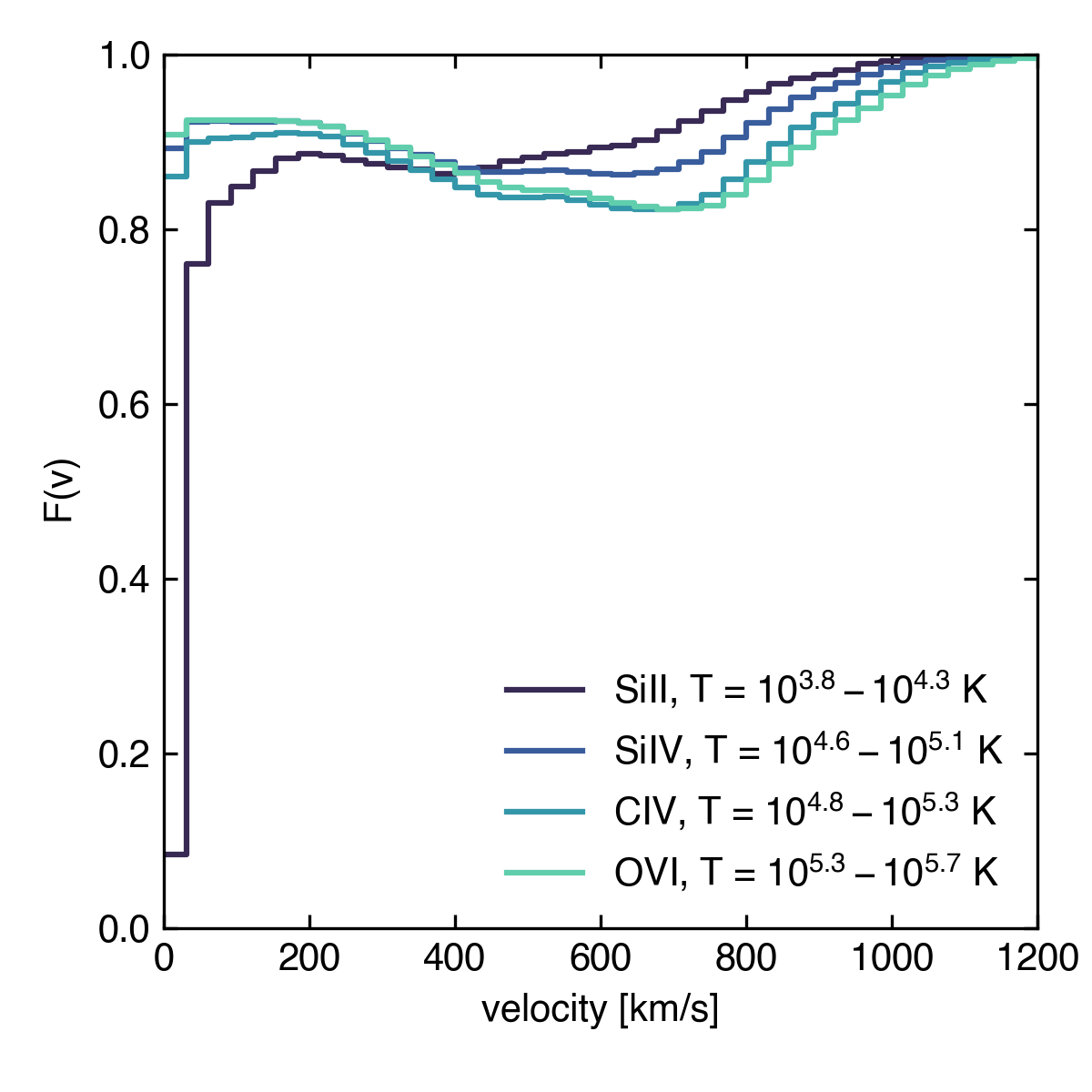}
    \caption{Mock ``down-the-barrel" absorption spectra after 30 Myr of feedback for the distributed feedback model (CGOLS~V, left) versus the central burst model (CGOLS~IV, right).}
    \label{fig:synthetic_absorption}
\end{figure*}

More quantitatively, we can estimate the covering fraction of Mg as a function of column density for our models. The distributed simulation has a covering fraction of 96\% at $N_\mathrm{MgII} > 10^{13}\,\mathrm{cm}^{-2}$ out to $10\kpc$, and 40\% at $N_\mathrm{MgII} > 10^{15}\,\mathrm{cm}^{-2}$. The analogous values are 61\% and 14\% for CGOLS IV. In addition, there is some evidence of a biconical structure in the cool gas, particularly for the CGOLS V model at $N_\mathrm{MgII} > 10^{15}\,\mathrm{cm}^{-2}$. We will investigate this azimuthal dependence further in future work using simulation volumes which extend to $20\kpc$ in all directions.

\subsection{Mock absorption profiles}

We additionally investigate mock ``down-the-barrel" absorption line profiles for several commonly observed ions. Following the procedure outlined in \cite{DelaCruz21}, we determine number densities for various elements in the simulation using solar abundances from \cite{Grevesse10} and assuming solar metallicity for all gas in the simulation. We then further calculate number densities per ion by assuming that each ion exists only in the temperature range specified by its full-width-half-maximum peak in collionsional ionization equilibrium (CIE) \citep[e.g.][Fig. 4]{Tumlinson17}. We do \textit{not} include the effects of photoionization in this modeling, which may contribute substantially for the lower potential ions considered here. We then calculate normalized fluxes as a function of vertical outflow velocity, starting at the disk midplane and integrating along the $z$-axis for each $(x,y)$ line-of-sight. We then average all $(x,y)$ sightlines within the central 5.0 kpc of the box, in order to better compare between the two outflow models, and assume a uniform background light source within that radius. Additionally, we only generate sightlines using the top half of the simulation domain to reproduce the effect of a midplane light source.

The resulting absorption lines are shown in Figure \ref{fig:synthetic_absorption} for the distributed cluster simulation (left) and the central cluster simulation (right) for several common ions. A few features are common to both. First, lower ionization potential lines (corresponding to lower temperature ranges in our model) have systematically lower velocities. Additionally, the lowest ionization lines are the deepest, indicating the highest column densities along individual sightlines. As stated above, our method for generating these synthetic spectra assumes a uniform background light source across the 5 kpc central region, and the profiles for all the $(x, y)$ sightlines are averaged in order to produce the total absorption line profile. Most lines-of-sight for the low ions are either saturated or zero at at given velocity, so the overall depth of these lines is set primarily by the covering fraction. For OVI, in contrast, many lines of sight have low optical depth, and the overall covering fraction is larger (see the discussion in \cite{DelaCruz21} for further details on this effect).

\begin{figure*}[ht]
    \centering
    \includegraphics[width=0.43\linewidth]{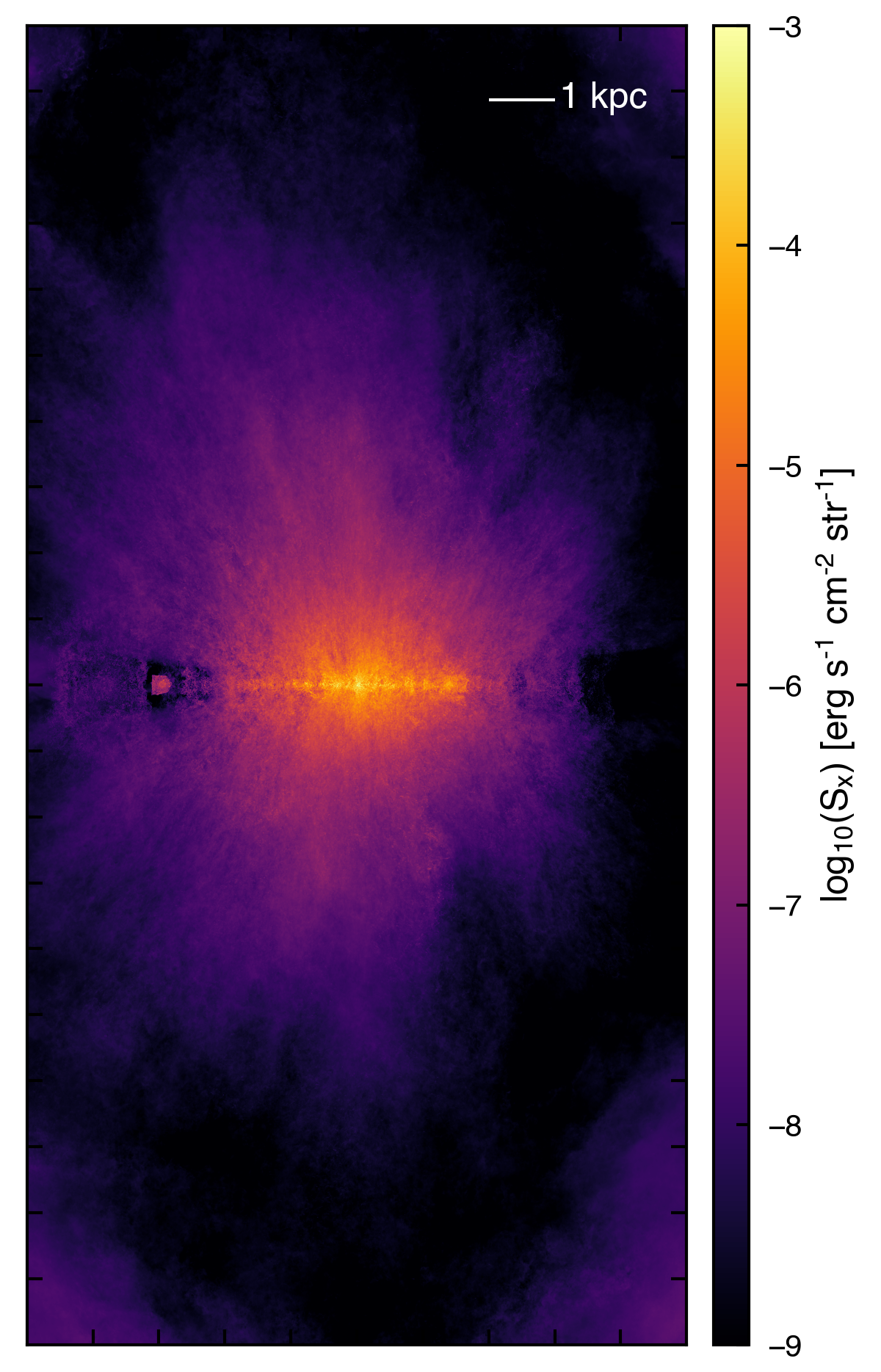}
    \includegraphics[width=0.43\linewidth]{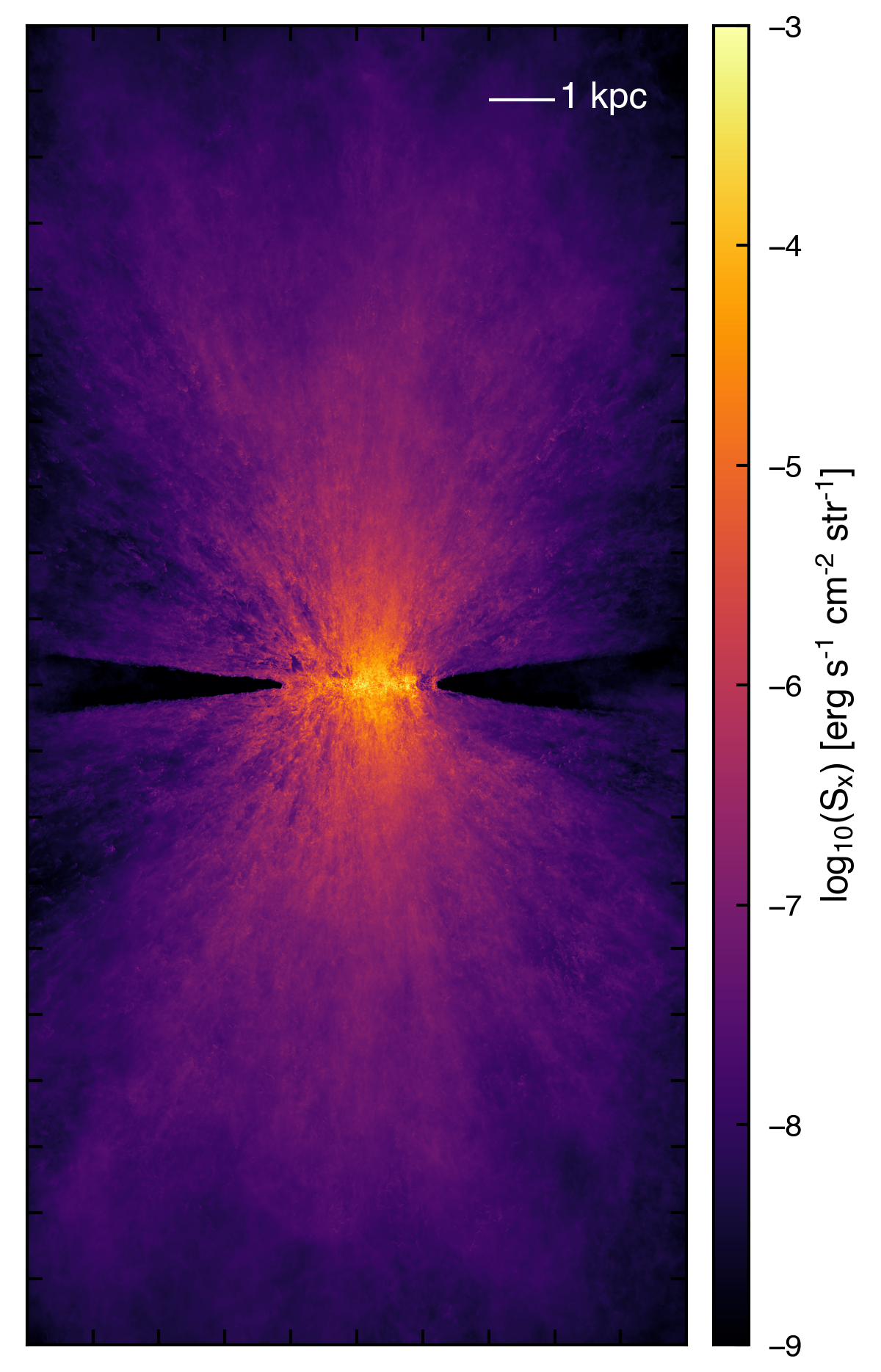}
    \caption{Soft X-ray surface brightness maps for the distributed (left) and central (right) burst models.}
    \label{fig:xrays}
\end{figure*}

There are, however, some differences in the shape of the line profiles between the central and distributed bursts. The distributed model has less uniform line profiles, with no clear single minimum for most ions. The central burst, by contrast, has two primary troughs, one at 0 velocity that corresponds to rotating gas in the galaxy disk, and one at higher velocity that increases in velocity with ionization potential. This is consistent with small-scale simulations which show that higher temperature gas in winds is moving more quickly \citep{Schneider17}. At any given velocity, the lines are significantly deeper in the distributed model, which is consistent with our findings in Section \ref{sec:results} that the distributed model has more cool gas at low $z$ and larger $R_\mathrm{cyl}$. A visual analysis of Figures \ref{fig:streamlines} and \ref{fig:mgII_maps} also indicates that there is substantially more cool gas at low heights and intermediate radii in the distributed feedback model, which can explain the extra absorption.

We emphasize that these mock absorption spectra should not be interpreted as directly comparable to observed spectra given the lack of contributions from photoionized gas, which may dominate the low ions at low $z$. Generating such spectra, including self-consistent contributions from the local ionizing background sources in the disk, is a topic of future work.

\subsection{Soft X-ray Surface Brightness}

Many nearby star-forming galaxies have been mapped in soft X-rays by the \textit{Chandra} observatory \citep{Strickland04b, Li13}. These photons are emitted by hot gas and are one of the few probes of the hot phase of galactic winds. Thus we conclude our Section on mock observables with an estimate of the soft X-ray surface brightness from both models.

Figure \ref{fig:xrays} shows surface brightness maps for CGOLS V and CGOLS IV. These maps were made following the procedure outlined in \cite{Schneider18a}. In brief, we use a temperature cut to identify gas cells with temperatures in the range $0.2 - 2.0\,\mathrm{eV}$, then estimate the emission from these cells using the same CIE cooling curve that was used in the simulations. We integrate the emission along each $(x,z)$ position to create the maps shown. Both models show distributed soft X-ray emission well in excess of the predictions from an adiabatically-expanding hot wind model, demonstrating the importance of cloud-wind interactions in generating the X-ray emission. The total X-ray luminosity for both simulations is comparable: the central burst model has a total integrated luminosity of $L_X = 1.4\times 10^{40}\,\mathrm{erg}\,\mathrm{s}^{-1}$, while for the distributed model we calculate $L_X = 1.1\times 10^{40}\,\mathrm{erg}\,\mathrm{s}^{-1}$. This reflects the fact that CGOLS IV contains slightly more hot gas, though overall the profiles are similar. Both simulations also compare favorably with the diffuse soft X-ray emission measured for M82, which is of order $L_X = 10^{40}\,\mathrm{erg}\,\mathrm{s}^{-1}$ \citep{Li13}.

\section{Discussion}

\subsection{Comparison to Simulations}\label{sec:sim_compare}

We now turn to a discussion these models in the context of other studies. In general, a number of numerical simulations in recent years have demonstrated that resolved supernova feedback can generate multiphase outflows with a structure similar to that seen in CGOLS -- namely a hot, volume-filling outflow with an embedded spectrum of cool clouds \citep[e.g.][]{Creasey15, Tanner16, Martizzi16, Li17, Fielding18, Emerick19, Hu19, Armillotta19, Martizzi20, Kim20a, Steinwandel22, Rathjen23, Vijayan23}. These range from simulations of $\sim \kpc$ patches of the ISM, or so-called ``tall box" simulations, to isolated galaxy simulations and the highest resolution cosmological zooms. In this Section, we focus our comparison on studies that measured outflow rates and loading factors, as that is the most straightforward.

\cite{Li20} found that for a variety of simulations, the energy loading is dominated by the hot phase, with values ranging from a few percent to $\eta_E = 0.3$ for all the considered simulations, which covered a broad range of star formation surface densities, $\Sigma_{SFR} \sim 5\times10^{-5} \Msun \kpc^{-2} - 1 \Msun \kpc^{-2}$. Our results are in line with these values for CGOLS V, with a measured $\eta_E \approx 0.1$ at $r = 5\kpc$ for an average $\Sigma_{SFR} \sim 3\times10^{-1} \Msun \kpc^{-2}$, although if we measure $\eta_E$ at $1\kpc$ (which is more similar to the heights used in most tall box simulations), we recover a slightly higher value of $\eta_E \approx 0.15$. By contrast, our measured value of $\eta_E = 0.5$ for CGOLS IV is considerably higher than any found in the simulations above, but we note that this centrally-concentrated burst also corresponds to a significantly higher $\Sigma_{SFR} \sim 6 \Msun \kpc^{-2}$ than any of the other simulations, and the suite of simulations by \cite{Kim20a} does demonstrate a positive trend between $\eta_E$ and $\Sigma_{SFR}$.

Only one simulation, \cite{Emerick19} found a comparable energy loading between the hot and cool phases, and that was for the lowest considered $\Sigma_{SFR} \approx 0.5\times10^{-5} \Msun \kpc^{-2}$. In this sense, our result that the hot and cool phase carry a similar amount of energy in the CGOLS V simulation is an outlier. However, we note that most of the simulations we are comparing against here measure loading factors at heights significantly less than 5 kpc due to limits on domain size, and at smaller radii, the CGOLS V model does have an energy loading dominated by the hot phase. All of those that extend to larger radii use an adaptive resolution, which inhibits mixing between phases. Thus, it is possible that a phase transition is occurring in our simulation as mass is carried out and the energy in the hot phase continues to be drained by mixing with the cool clouds, which is not captured in smaller boxes or at lower resolutions.

In Section \ref{sec:fluxes} we demonstrated that the total mass loading in the the central burst model (CGOLS IV) and the distributed cluster simulation (CGOLS IV) is similar. When measured across the full sphere at 5 kpc, we recover approximately $\eta_m = 0.2$ in both models. The fraction of the mass carried by the hot versus the cool phase differs, however, with the majority of the mass flux carried by the hot phase for CGOLS IV, and by the cool phase for CGOLS V. \cite{Kim20a} measure mass-loading factors for simulations with a range of star formation rate surface densities. In all of their models, the majority of the mass was carried by the cool phase, with increasing ratios of $\eta_{m, cool}$ to $\eta_{m, hot}$ as $\Sigma_{SFR}$ decreased. In general, their measured values of $\eta_m$ were significantly higher than those we quote here, however their measurements were made at the scale height of the disk, and as we can see from Figure \ref{fig:full_fluxes}, the mass-loading factor drops significantly in our model as a function of distance, a behavior that is consistent with measurements at different scale heights in simulations with planar geometry \citep{Martizzi16, Kim20a}.

Given this context, we also compare our mass loading factors to a zoom-in model from the FIRE 2 simulation suite measured at a comparable spherical radius \citep{Pandya21}. For an M82-mass galaxy at $z = 0$, they find $\eta_m$ values of approximately 0.25, in good agreement with our models. However, they find that for all cases when $\eta_{m, tot} < 1$, the hot phase dominates the mass-loading rate, whereas we see two different modes between our more and less concentrated $\Sigma_{SFR}$. Although isolated galaxy simulations at comparable resolution to CGOLS have been run \citep{Emerick19, Hu19, Steinwandel22}, they all focused on dwarf galaxies with considerably smaller potentials and much lower $\Sigma_{SFR}$ than the CGOLS fiducial model. These simulations tend to find higher $\eta_m \sim 1$, which is consistent with our work, assuming the negative trends between mass loading and $\Sigma_{SFR}$ are correct \citep{Kim20b, Steinwandel22}. We tend to find much higher energy loading than these models.

\subsection{Comparison to Observations}\label{sec:obs_compare}

Many observational studies of outflows have been conducted in the past several decades, and it would be impossible to do justice to a full comparison here. While outflows have been observed with a wide variety of instruments and telescopes and in a variety of phases, the largest samples exist for the cool phase ($T \sim 10^4\,\mathrm{K}$), particularly for local galaxies. Thus, we will focus our comparison on these data, though we note that other phases including X-ray probed hot gas and cold neutral and molecular gas may also prove equally constraining for theoretical models \citep[e.g.][]{Veilleux20, Lopez20, Nguyen21}.

Using UV absorption line data from the CLASSY survey \citep{Berg22}, \cite{Xu22} compile a sample of 50 nearby starburst galaxies covering a range of stellar masses and star formation rates. ``Down-the-barrel" spectra of several low ionization lines at a variety of inclination angles allow them to directly measure a variety of useful parameters, including outflow velocities and cool gas covering fractions, and indirectly estimate additional features like mass and metal outflow rates. In general, we find good agreement between the CGOLS V model and these data. In particular, their best-fit scaling relations imply that an M82-like galaxy would have a mass-loading factor $\eta_m \sim 0.3 - 0.6$, which is entirely consistent with our model, in which $\eta_m$ peaks at 0.5 and falls to 0.25 at the edge of the simulation domain. Similarly, they estimate a cool-phase energy-loading factor of 5\% for an M82-like galaxy, which is comparable to our measured value of 3\%.

We can also directly compare the outflow velocities of the cool phase, noting that their nomenclature uses the median velocity of the fitted absorption line's FWHM as the ``outflow velocity". Assuming a similar radius as the observations, we see that the median value for our $v_\mathrm{r,cool}$ ranges from $\approx 200\kms$ at 4 kpc to $\approx 400\kms$ at 8kpc. The estimated range of outflow velocities in \cite{Xu22} for an M82-like galaxy is $260 - 360\kms$, again an excellent agreement. We note that these values are also a better fit to the data than the CGOLS IV model, which had lower cool gas mass-loading and higher velocities.

\section{Conclusions}\label{sec:conclusions}

We have presented an analysis of CGOLS V, the fifth simulation in the Cholla Galactic OutfLow Simulations suite. This simulation models a $10^{10}\,M_\odot$ starburst galaxy with thermal supernova feedback injected in individual clusters distributed in a pattern following the gas surface density. Our primary conclusions from this analysis include:
\begin{enumerate}
    \item Resolved supernova feedback generates a multiphase outflow with gas at a large range of densities and temperatures (see Figure \ref{fig:slices}).
    \item The number density, pressure, and temperature in the hot phase of the outflow falls off with radius at rates that are flatter than predicted by analytic models of adiabatic expansion (see Figure \ref{fig:profiles_hot}).
    \item The cool phase is distributed in a population of clouds, with a gas number density that falls off with distance as $r^{-2}$ and slowly rising velocities (see Figure \ref{fig:profiles_cool}).
    \item The hot and cool phases of the outflow are not in pressure equilibrium (see Figure \ref{fig:p_compare}).
    \item Total mass outflow rates are comparable to a simulation with the same star formation rate but more centrally-concentrated stellar feedback (CGOLS~IV), but $6\times$ more mass exists in the cool phase of the outflow in the more distributed model (CGOLS V) (see Figure \ref{fig:phase_plots}).
    \item Energy outflow rates in the more distributed model are lower than in the centrally-concentrated burst, and energy-loading between the hot and cool phases is comparable (see Figure \ref{fig:fluxes_compared}).
    \item More distributed star formation results in significantly higher covering fractions of cool gas at all radii (see Figure \ref{fig:mgII_maps}).
    \item Mass, momentum, and energy outflow rates in the cool phase for the distributed model are consistent with observations of local starburst galaxies.
\end{enumerate}

\begin{acknowledgements}
EES thanks Kate Rubin for many helpful discussions that have enriched this work. This research used resources of the Oak Ridge Leadership Computing Facility, which is a DOE Office of Science User Facility supported under Contract DE-AC05-00OR22725, using Titan allocation INCITE AST125 and Summit allocation CAAR CSC380. This research was supported in part by the University of Pittsburgh Center for Research Computing, RRID:SCR\_022735, through the resources provided. Specifically, this work used the H2P cluster, which is supported by NSF award number OAC-2117681. E.E.S. acknowledges support from NASA TCAN grant 80NSSC21K0271, NASA ATP grant 80NSSC22K0720, StScI grant HST-AR-16633.001-A, and the David and Lucile Packard Foundation (grant no. 2022-74680).
\end{acknowledgements}


\software{\texttt{Cholla} \cite{Schneider15}, \texttt{numpy} \citep{VanDerWalt11}, \texttt{matplotlib} \citep{Hunter07},  \texttt{hdf5} \citep{hdf5}; Cloudy \citep{Ferland13}, NVIDIA IndeX}

\bibliography{all}{}
\bibliographystyle{aasjournal}

\end{document}